\documentclass[
 reprint,
 amsmath,amssymb,
 aps,
]{revtex4-2}

\usepackage{graphicx}
\usepackage{dcolumn}
\usepackage{bm}
\usepackage{amsmath}
\usepackage{esint}
\usepackage[dvipsnames]{xcolor}

\usepackage[mathlines]{lineno}

\begin{document}

\preprint{AAPM/123-QED}

\title[Sample title]{Implementing Josephson Junction spectroscopy in a scanning tunneling microscope}

\author{Margaret A. Fortman} 
\affiliation{University of Wisconsin-Madison, Department of Physics, 1150 University Ave., Madison, Wisconsin 53706, USA}
\author{David C. Harrison}
\affiliation{University of Wisconsin-Madison, Department of Physics, 1150 University Ave., Madison, Wisconsin 53706, USA}
\author{Zachary J. Krebs} 
\affiliation{University of Wisconsin-Madison, Department of Physics, 1150 University Ave., Madison, Wisconsin 53706, USA}
\author{Ramiro H. Rodriguez}
\affiliation{Quantronics Group, Université Paris-Saclay, CEA, CNRS, SPEC
91191 Gif-sur-Yvette Cedex, France}
\author{Sangjun Han}
\affiliation{Korea Advanced Institute of Science and Technology, School of Electrical Engineering, 291 Daehak-ro, Yuseong-gu, Daejeon 34141, Korea}
\author{Min Seok Jang}
\affiliation{Korea Advanced Institute of Science and Technology, School of Electrical Engineering, 291 Daehak-ro, Yuseong-gu, Daejeon 34141, Korea}
\author{\c Ca\u glar \"O. Girit}
\affiliation{Quantronics Group, Université Paris-Saclay, CEA, CNRS, SPEC
91191 Gif-sur-Yvette Cedex, France}
\author{Robert McDermott}
\affiliation{University of Wisconsin-Madison, Department of Physics, 1150 University Ave., Madison, Wisconsin 53706, USA}
\author{Victor W. Brar}
\affiliation{University of Wisconsin-Madison, Department of Physics, 1150 University Ave., Madison, Wisconsin 53706, USA}
\email{vbrar@wisc.edu}

\begin{abstract}
Josephson junction spectroscopy is a powerful local microwave spectroscopy technique that has promising potential as a diagnostic tool to probe the microscopic origins of noise in superconducting qubits. We present advancements toward realizing Josephson junction spectroscopy in a scanning geometry, where the Josephson junction is formed between a superconducting sample and a high capacitance superconducting STM tip. Data from planar Nb-based Josephson junction devices first demonstrate the benefits of including a high capacitance shunt across the junction, which decreases linewidth and improves performance at elevated temperatures. We show how an equivalent circuit can be implemented by utilizing a planarized STM tip with local prominences, which are fabricated via electron beam lithography and reactive ion etching, followed by coating with a superconducting layer. Differential conductance measurements on a superconducting NbN surface demonstrate the ability of these high capacitance tips to decrease both thermal noise and P(E)-broadening in comparison to typical wire tips.

\end{abstract}

\maketitle

Impurity-driven interactions on superconductors have attracted significant attention recently due to their detrimental effects on qubits. Defects that form spurious two-level systems (TLS) are a major source of decoherence \cite{Towardsunderstanding2019}. TLS defects have been observed by a number of techniques using on-chip devices, such as electric field tuning of defects \cite{Landau2014, ProjectedDipole2016, Laser2016, Transmissionline2017, Dynamical2019, Quantumsensors2021, Quantifying2021, Experimentallyrevealing2022, Probinghundreds2022, Enhancing2023, redshift2024, Giant2024} and  strain-spectroscopy \cite{StrainTuning2012, Observation2015, Probingdefect2022}.
These experiments have also attempted to localize the defects by studying surface participation ratios \cite{Wenner2011, Sandberg2013, surfacepart2015, Bulkandsurfaceloss2016, Dial2016, InvestigatingSurface2017, AnalysisandMitigation2018,  DeterminingInterface2019, Verjauw2022, Wang2022, Eun2023} and applying electric fields to samples to distinguish defects in tunnel barriers from those at electrode interfaces \cite{efieldspec2019, Resolving2020}. Developing new diagnostic tools to understand superconducting surface impurities is crucial for the future of quantum superconducting circuity as well as probing new emergent states in condensed matter, such as Majorana fermions \cite{majorana1, majorana2, majorana3, majorana4, majorana5}. An ideal probe of surface defects would have both atomic resolution and the ability to measure the energetic structure of a specific defect and correlate that structure with the surrounding physical environment. Scanned probe microscopy (SPM) techniques can provide such information, but typical methods cannot resolve microwave absorption --- which is the source of TLS loss --- and the energetic resolution in electron tunneling measurements is limited by thermal broadening.  In principle, these challenges could be overcome using scanning microwave spectroscopy methods, where $\sim$ GHz microwaves are sent to and/or received from an SPM junction \cite{RFSTM2007, MAllen2023}, however such techniques require complex impedance matching and have low spatial resolution.  
\begin{figure}[b]
    \includegraphics[width=3.4in,keepaspectratio]{
    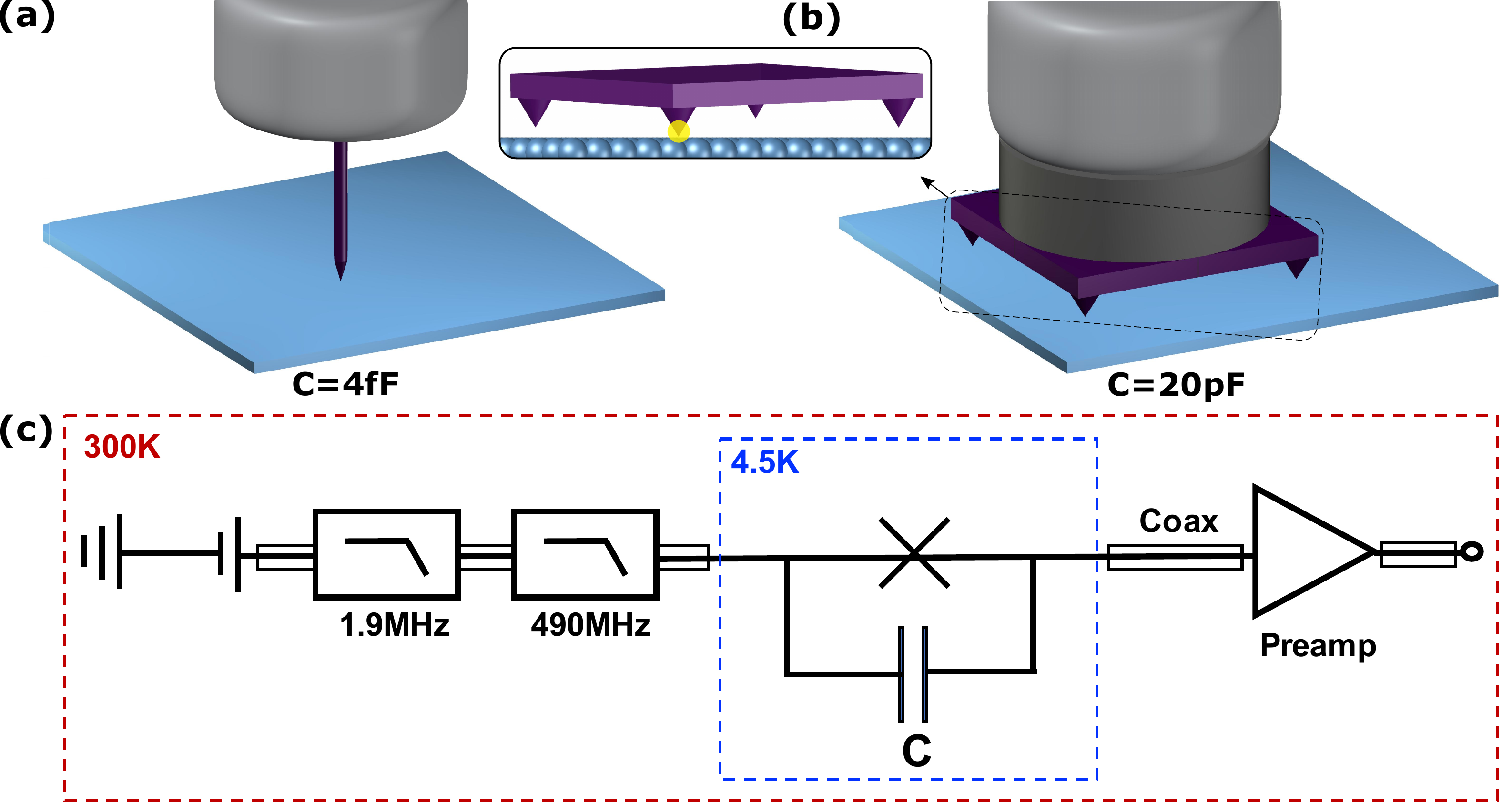}
    \caption{\label{fig1} (a) Representation of low capacitance STM tip consisting of NbN-coated wire tip on a superconducting NbN sample, creating a 4fF tunnel junction. (b) Representation of high capacitance Josephson STM tip on superconducting NbN sample, creating a 20pF tunnel junction. Inset shows one of the four tips tunneling into substrate. (c) Circuit diagram of STM measurement setup. Lines at 4.5K (not denoted coax) are bare copper wires. }
\end{figure}

\begin{figure*}
    \centering
    \includegraphics[width=6.8in,keepaspectratio]{
    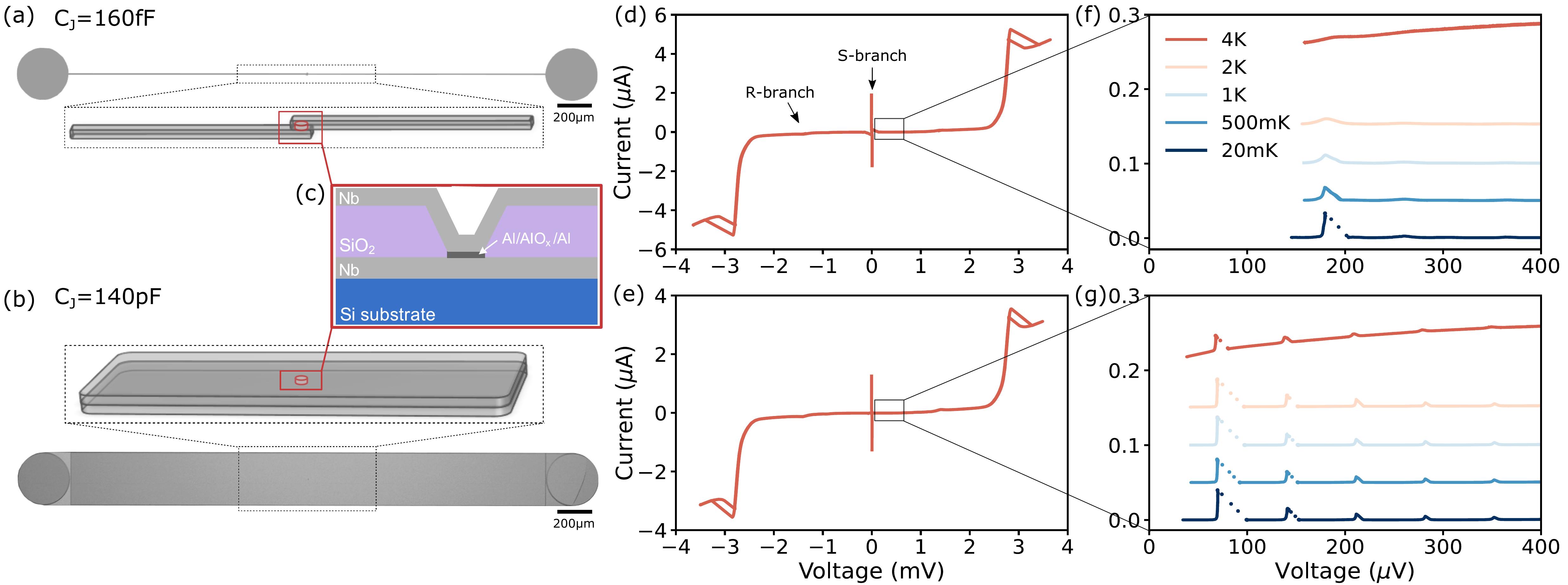}
    \caption{ Schematic of overlapping Nb planes creating the (a) low capacitance ($\sim$160fF) and (b) high capacitance (140pF) planar junctions. (c) Cross section of devices showing schematic of fabricated Nb/Al-oxide/Nb junctions. (d-e) Full I-V spectra at 20mK for each device. (f-g) Low voltage I-V spectra at increasing temperatures from 20mK to 4K.}
    \label{junctions}
\end{figure*}

\begin{figure}[b]
    \centering
    \includegraphics[width=3.4in,keepaspectratio]{
    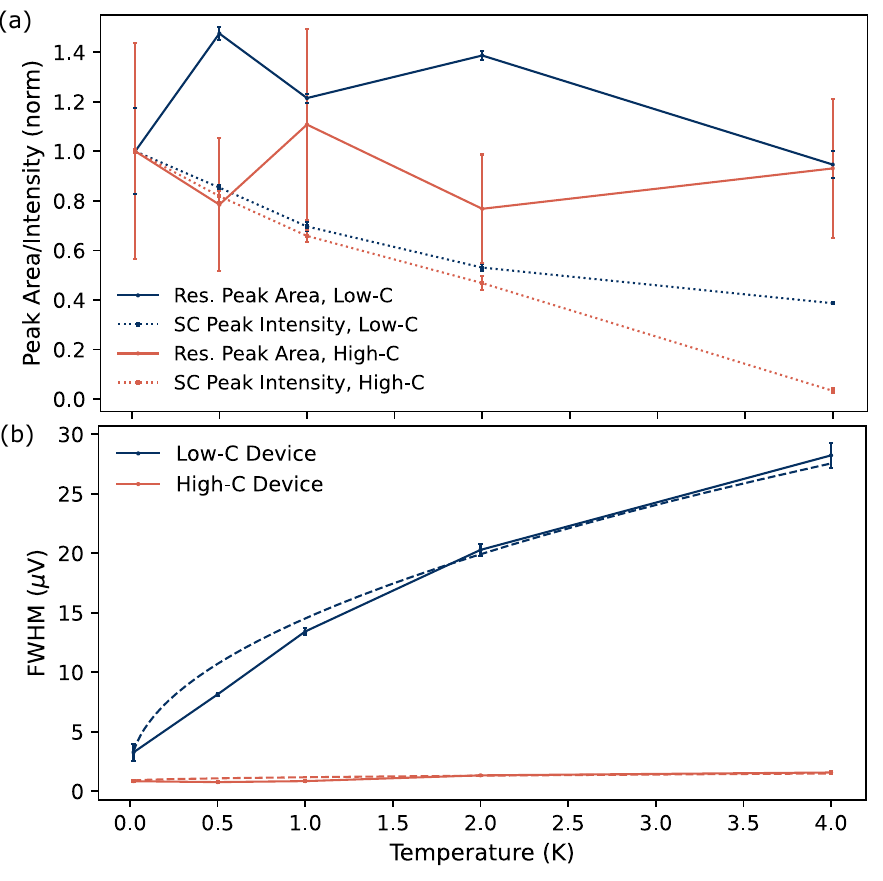}
    \caption{(a) Resonant peak area and supercurrent peak intensity vs temperature for both low capacitance devices and high capacitance devices. (b) FWHM of first resonance peak vs temperature with fit.}
    \label{pkarea&fwhm}
\end{figure}

In this work we discuss how an alternative microwave spectroscopy method, Josephson junction spectroscopy (JJS), could potentially be implemented in a scanning probe geometry.  JJS is unique in that it produces tunable microwaves over a large bandwidth using only DC voltages. At non-zero voltage bias a junction undergoes AC Josephson oscillations which when coupled to a resonant mode results in a DC current. This allows a Josephson junction to act as both an emitter and detector of microwaves that can be used to probe nearby defects, impurities, and antenna resonances. Devices have already been implemented to perform Josephson junction spectroscopy on-chip to detect a variety of transitions \cite{Griesmar}, including zero-field nuclear magnetic resonances \cite{Silver1967} and Andreev states  \cite{Bretheau2013, Girit2}. The advantages of using this technique in a scanning tunneling microscope (STM) are three-fold: no complex wiring is required, emitted microwaves have linewidth determined by $\sqrt{k_BT/C}$ rather than the $\sim k_B T$ resolution of typical STM measurements, and it adds the potential to probe electronic and photonic transitions simultaneously.

Here, we present an advancement in combining the benefits of JJS with STM, which would enable microwave absorption spectroscopy with nanometer resolution. We note that Josephson STM itself -- where a superconducting tip is used to probe a superconducting sample -- has been implemented in a number of experiments.  This was first pioneered by Dynes \cite{Dynes} and more recently has allowed, for example, measurements of the pair-density of unconventional superconductors \cite{Cho2019,Hamidian2016,randeria2016, proslier2006, rodrigo2006, bergeal2008} and enhanced STS resolution of superconducting samples \cite{Pan1998, Franke2011, Franke2015}. Here we focus on the possibility of probing and engineering the microwave emission generated at the tip-sample Josephson junction. This has also been explored by other groups, and there are two known challenges that arise: the STM tip acts like a monopole antenna which creates broad resonances that absorb radiation \cite{Jack2015, Ast2016,roychowdhury2015,bastiaans2019, randeria2016}; the STM tip has very low capacitance (on the order of 1fF) which cannot filter microwave noise, and produces a large emission linewidth for a given temperature\cite{EnvironmentAssisted2010,roychowdhury2015,Jack2015, Ast2016}.  In order to partially address those issues, previous measurements that explored JJS in an STM have been conducted at temperatures below 1K,  which allows for narrow linewidths even for low capacitance junctions. This paper presents an alternative approach towards JJS-STM by engineering an STM tip that, unlike a conventional wire tip, 
contains spectral gaps in its environmental energy spectrum (Supplementary Figure S1) and has a capacitance four orders of magnitude larger than a conventional metal wire tip (Figure \ref{fig1}), which theoretically would allow for JJS-STM to be effectively implemented at elevated temperatures. In this paper, we first demonstrate the benefits of implementing JJS with a large cross junction capacitance by testing on-chip devices and then present the fabrication procedure and performance of a high capacitance Josephson STM tip.

To demonstrate the importance of this large cross-junction capacitance, we first start by testing on-chip Nb-based Josephson junction devices. We tested two different geometries.  The first (Figure \ref{junctions}(a)) consisted of a Josephson junction contacted by two 1.35mm long Nb contacts that had a small amount of overlap across the junction creating a $\sim$160fF cross-junction capacitance.  The second (Figure \ref{junctions}(b)) was formed by a Josephson junction contacted-by and embedded-between two 0.3mm by 2.7mm Nb planes that created a 140pF cross-junction capacitance. The junctions of these devices are 2$\mu$m in diameter and were fabricated using a Nb/Al/AlO$_x$/Al/Nb structure (as shown in Figure \ref{junctions}(c)), which leverages both the high critical temperature of Nb and the high quality of aluminum native oxide (AlO$_x$) to create the insulating barrier.

Measurements of these on-chip JJS devices were performed in a closed-cycle dilution refrigerator at a base temperature of about 20mK. All lines inside the fridge are heavily filtered flat cables. The junction is DC biased and cold filtered with $\sim$100nF parallel capacitance. The output current is measured with a low noise preamplifier (Supplementary Figure S2). The characteristic I-V curves for both devices (Figure \ref{junctions}(d-e)) show a zero-bias supercurrent peak (`S-branch') within a low-conductance superconducting bandgap where the current is minimal (`R-Branch').  Within the gap, sharp resonances can be observed (Figure \ref{junctions}(f-g)) with fundamental frequencies of $V_{\textrm{low-C}}=90\mu V$ and $V_{\textrm{high-C}}=69\mu V$ for the low and high capacitance device, respectively.  The energies of these modes match calculations of the standing wave resonances in each device that couple to the microwave emission from the junction (see Supplementary Figures S3 and S4). 
Only the left side of the resonance peaks are visible due to a biasing instability (negative differential resistance) on the right side of the peaks.

A feature of Josephson junction spectroscopy is that it operates in the non-zero voltage region of the current-voltage characteristic (R-branch) and not on the supercurrent peak at zero voltage (S-branch).  As compared to the supercurrent peak, which may easily be smeared by current fluctuations and phase diffusion, peaks on the R-branch have greater noise immunity.  Voltage fluctuations are reduced as the AC Josephson oscillations phase lock to the coupled resonant mode \cite{likharev1985}.  This effect is shown experimentally in Figure \ref{junctions}(f-g), where spectra were taken for each device at temperatures increasing from 20mK to 4K. Analyzing peak area in Figure \ref{pkarea&fwhm}(a) shows the resonance peaks in the R-branch are indeed minimally effected by increasing temperature, while the supercurrent peak in the S-branch is suppressed with increasing temperature for both devices. 

\begin{figure}[h]
    \centering
    \includegraphics[width=3.4in,keepaspectratio]{
    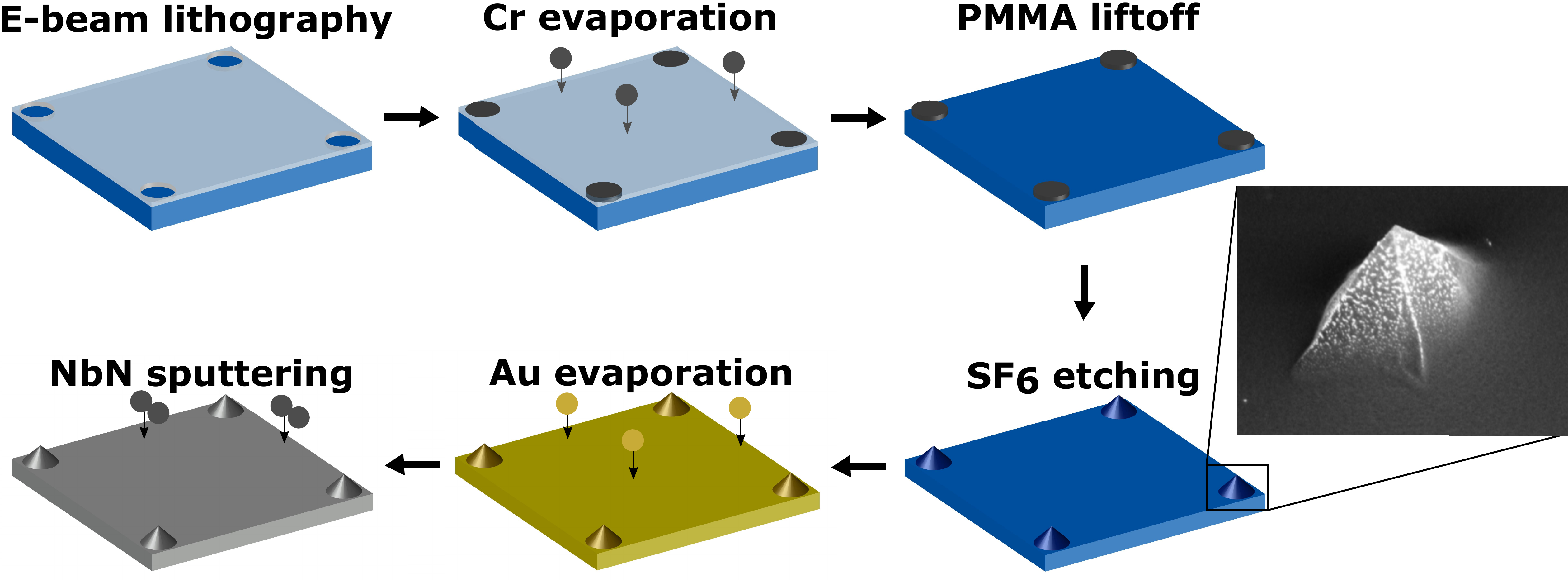}
    \caption{Fabrication procedure of high capacitance Josephson STM tip. Electron beam lithography is used to pattern Cr masks at each corner of the substrate. An ICP etch is used to cut the tip from the substrate to create a suspended apex. The entire device is coated in Au and finally NbN is deposited to create superconducting tips. SEM image of atomically sharp tip protruding from substrate. }
    \label{fab}
\end{figure}

The benefit of large cross-junction capacitance is demonstrated in Figure \ref{pkarea&fwhm}(b) comparing the linewidth (FWHM) of the first resonance peak for each device with increasing temperature. At high temperatures, the linewidth for the low-capacitance device significantly increases, while it remains relatively unchanged for the high-capacitance device. We expect the linewidth to depend on both capacitance and temperature as $\sqrt{k_B T/C}$. 
The data fits well to the function $a\cdot(\Gamma_T+\Gamma_i)$ where $a$ is a fitting parameter ($a_{\textrm{high-C}}=1$ and $a_{\textrm{low-C}}=1.4$), $\Gamma_i$ is the intrinsic linewidth which accounts for the inherent loss in the resonance ($\Gamma_{i,\textrm{high-C}}=0.86\mu$eV and $\Gamma_{i,\textrm{low-C}}=1.5\mu$eV), and $\Gamma_T=\sqrt{k_BT/C}$. 
The good agreement observed between our data and a simple model of linewidth defined by temperature and capacitance demonstrates the straightforward manner that capacitance controls linewidth in JJS.  These results also show that for devices with large cross-junction capacitance, JJS can be effective at elevated temperatures that are fractions of $T_c$.

\begin{figure}[h!]
    \centering
    \includegraphics[width=3.4in,keepaspectratio]{
    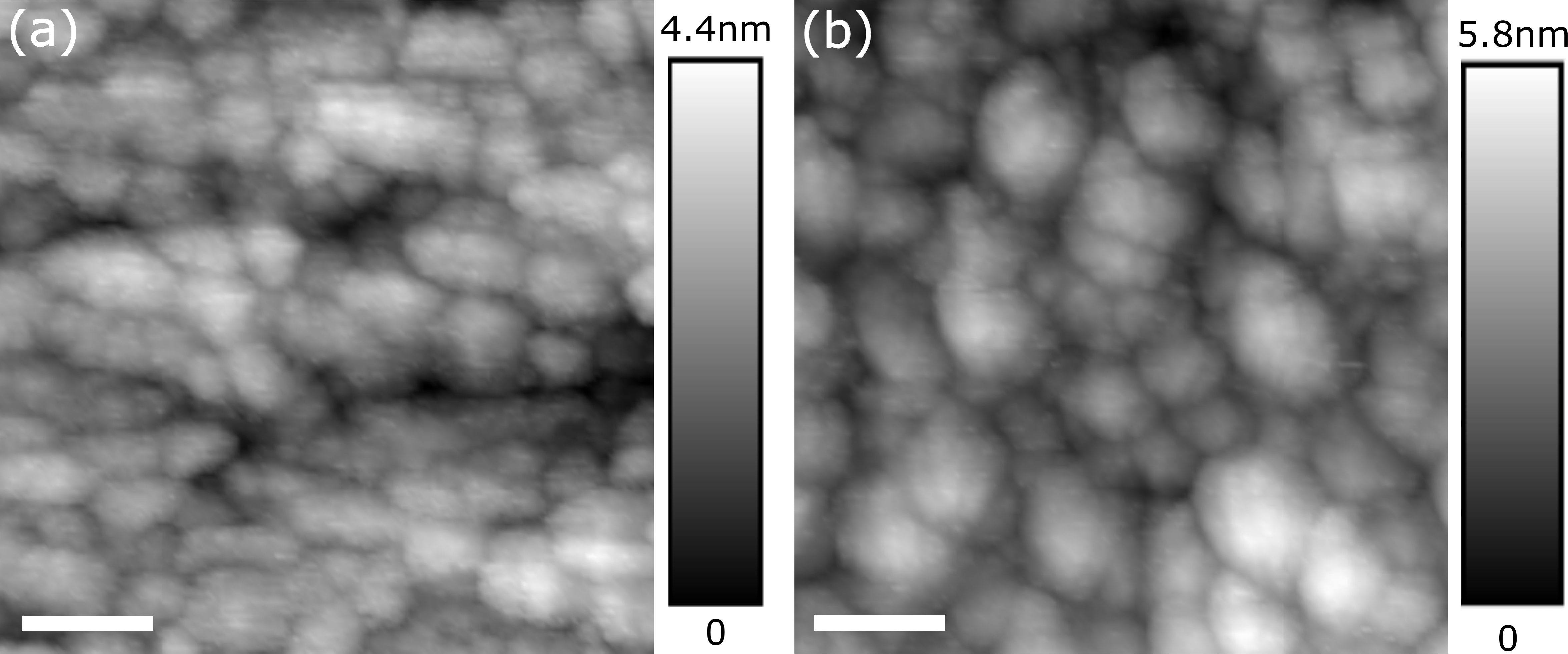}
    \caption{Topography of ALD-grown grains of NbN surface taken with (a) high capacitance Josephson STM tip and (b) low capacitance wire tip at 4.5K (V$_b$=-500mV, I=50pA). 10nm scale bar.}
    \label{topography}
\end{figure}

We seek to leverage this advantage of increased cross-junction capacitance by designing and testing a superconducting STM tip that forms a high capacitance Josephson junction when in contact with a superconducting surface.  The basic geometry of this tip is illustrated in Figure \ref{fig1}.  The typical wire tip is replaced by a flat superconducting surface with short, sharp protrusions from each corner.  The fabrication procedure for these tips is illustrated in Figure \ref{fab}. We start by spincoating an electrosensitive photoresist (PMMA 950 A4) on 5mm square chips of Si(100). Electron-beam lithography is used to create a mask through which 20nm of Cr is evaporated. After an acetone liftoff, four Cr discs remain, one at each corner of the substrate. These discs act as a mask in the subsequent etching step. Inductively coupled plasma (ICP) etching is used to selectively etch the Si with SF$_6$, which causes the pyramid shape to form under the Cr masks. With further etching the Si underneath is pinched-off and the mask is dropped. By stopping the etch at this point, we obtain an atomically sharp tip. A scanning electron microscope (SEM) image of one of these tips is shown in the inset of Figure \ref{fab}. The entire device is then covered with gold which remains conducting at low temperatures. Finally, the tips are coated in NbN by sputtering. Creating a tip at each corner of the device guarantees that one of them will be the closest contact to the sample that is scanned over and, therefore, is robust against any sample tilt. Fabricating the tip on a `large' ($\sim 25$mm$^2$) chip gives us a large capacitance (20pF) between the tip and sample compared to a normal STM-Josephson junction ($\sim$4fF) \cite{Ast2016}.

\begin{figure}[h!]
    \includegraphics[width=3.4in,keepaspectratio]{    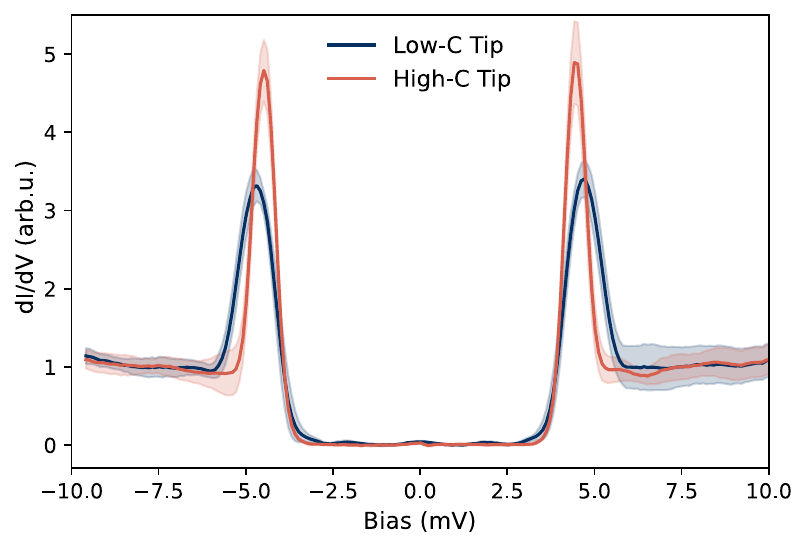}
    \caption{Measured dI/dV spectra of the superconducting densities of state for the low capacitance STM tip and the high capacitance Josephson STM tip on NbN sample. Plotted average and standard deviation over 13 spectra for each tip. Data taken at V$_b=-10$mV, I$_{\textrm{set}}=100$pA, lock-in V$_{\textrm{mod}}=200\mu$V peak to peak.}
    \label{dIdV}
\end{figure}

Next, we demonstrate the performance of the high capacitance Josephson STM tip and compare it to a conventional STM tip with low capacitance. To do so, we use the tips to measure an ALD-grown NbN substrate. These measurements are performed in an STM operating at 4.5K under UHV conditions. The NbN substrate is prepared in UHV by Ar ion sputtering and subsequent annealing for 1 hour at 300\textdegree C. The low capacitance STM tip is made from a Nb wire cut to a sharp apex and coated with NbN by sputtering. Both tips are loaded into UHV and annealed for 1 hour at 300\textdegree C (low capacitance tip) and 150\textdegree C (high capacitance tip). To demonstrate imaging capabilities, topography of the NbN surface taken with each tip is shown in Figure \ref{topography}. The ALD-grown NbN is observed by both tips to be rough, with a grainy structure and characteristic peak-to-peak roughness of approximately 5nm.  The observed difference in grain widths could be attributed to the microscopic shape of the tip apex or intrinsic variation of the NbN across the sample surface.  

\begin{figure}[h!]
    \centering
    \includegraphics[width=3.4in,keepaspectratio]{
    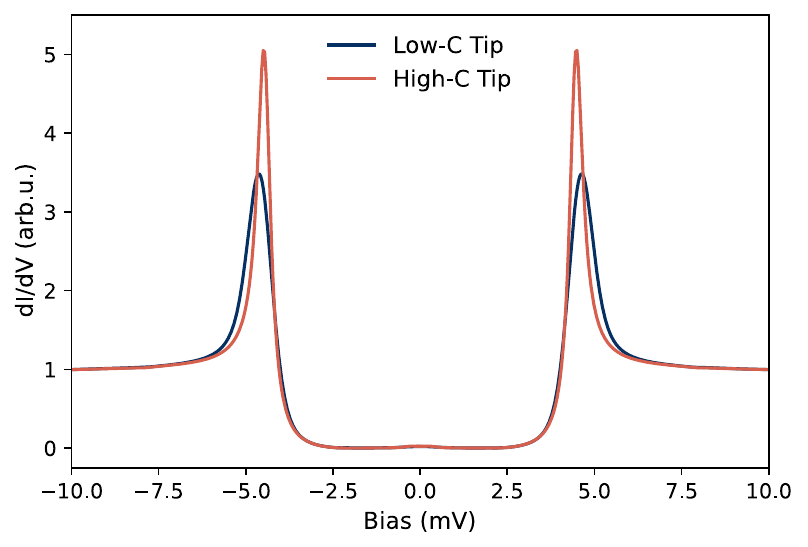}
    \caption{Calculated dI/dV spectra of superconducting densities of state for low  capacitance STM tip and high capacitance Josephson STM tip on NbN sample using P(E)-theory. }
    \label{PEdynes}
\end{figure}

The differential conductance dI/dV spectra measured with each tip as a function of bias voltage are shown in Figure \ref{dIdV}. Optimization of the filtering setup was done to attain the best spectra with highest energy resolution and conditions were then held constant across all data. Measurements were taken with the optimal setup of an in-series combination 490MHz and 1.9MHz low-pass filters on the bias line and electromagnetically-shielded Femto preamplifier on the current line (Figure \ref{fig1}(c)). As there is variation in spectra taken at different locations on the sample due to variation in NbN quality \cite{Brun}, we have taken an average over spectra at 13 random different positions on the sample within a 10x10nm area. Figure \ref{dIdV} shows tunneling spectra with superconducting gaps of width $2(\Delta_t+\Delta_s)\approx8.9$meV for the high capacitance Josephson STM tip and 9.1meV for the low capacitance STM tip, where $\Delta_s$ is slightly smaller than both values of $\Delta_t$ likely due to different NbN growth methods on tips and sample.  For both tips, a zero bias peak due to thermalized quasiparticles and weak Andreev bound states can be observed.  No supercurrent peak or R-Branch resonances are observed for either tip, and we associate this with the conventional wiring scheme of our STM, which lacks low temperature microwave filtering in our system near the sample.  This permits voltage noise at both the ground and bias lines of our sample.  To support this reasoning, an on-chip Nb JJS device was placed in the STM cryostat and measured with conventional STM wiring as well as with a filter mounted near the JJ (see Supplementary Fig. S5).   With the filter, both the S-Branch and standing wave resonances in the R-Branch were clearly observed, while they were not visible with the filter removed.  

Despite the lack of observable R-Branch resonances, the effect of the high capacitance tip on tunneling spectra can be observed in the coherence peaks at $(\Delta_t+\Delta_s)$.  In particular, the high capacitance Josephson STM tip has both taller and narrower coherence peaks in comparison to the low capacitance tip.  This difference can be associated with a difference in coupling of tunneling quasiparticles to the electromagnetic environment.  This process broadens the energy bandwidth of tunneling particles and therefore changes the spectroscopic resolution, with higher tip-sample capacitance leading to a higher energy resolution \cite{Ast2016}. In order to show this explicitly, we model the superconducting spectra for both tips using the framework of P(E)-theory, which models the energy exchange with the electromagnetic environment combined with the thermal noise across the junction capacitor \cite{Ast2016}. The tunneling current $I(V)$ from tip to sample is given by \cite{Devoret}:

\begin{widetext}
\begin{equation}
I(V) = \frac{1}{eR_T} \int_{-\infty}^{+\infty}\int_{-\infty}^{+\infty} dEdE'n_t(E)n_s(E'+eV) \left[ f(E)(1-f(E'+eV))P(E-E') -(1-f(E))f(E'+eV)P(E'-E) \right] 
\end{equation}
\end{widetext}
where $R_T$ is the tunneling resistance, $f(E)$ is the Fermi function, $n_t$ is the tip density of states, 
and $n_s$ is the sample density of states, both given by the Dynes equation \cite{DynesEq}:
\begin{equation}
    n_{t,s}(E)= \textrm{Re}\left[\frac{|E-i\Gamma|}{\sqrt{(E-i\Gamma)^2-\Delta_{t,s}^2}}\right],
\end{equation}
where $\Gamma$ is a material-dependent phenomenological parameter that we fit to 0.115meV . This value is of similar magnitude to previous measurements of NbN \cite{Brun}.

In our case of a very high impedance, the P(E) function is dominated by the thermal noise across the capacitance of the tunnel junction which takes the form \cite{Ingold1992}:
\begin{equation}
    P(E) = \frac{1}{\sqrt{4\pi E_ck_BT}}\textrm{exp}\left[ -\frac{(E-E_c)^2}{4E_ck_BT}\right]
\end{equation}
where $E_c$ is the charging energy $Q^2/2C_J$ with $Q=2e$ (charge of a Cooper pair). This function models the coherence peaks of both tips as we do not have in-gap resolution. The resulting dI/dV spectra for the capacitance of each tip is shown in Figure \ref{PEdynes}. We estimate the junction capacitance of the low-C tip to be 4fF which matches well with our measurements and is consistent with previous Josephson STM experiments \cite{Ast2016}. The dI/dV spectrum is sensitive to this value and the effect of varying the junction capacitance of the low-C tip can be seen in Supplementary Figure S6. The effect of larger capacitance on the voltage-dependent conductivity is observed as an increased height and decreased width of the coherence peaks, consistent with what was observed experimentally in Figure \ref{dIdV}.  We emphasize that this effect is due to the large capacitance decreasing the thermal voltage noise which results in P(E)-broadening of the spectra. From this P(E)-modeling, we can extract the intrinsic energy broadening of tunneling particles defined in Eq. (3), and find that the large capacitance tip exhibits a bandwidth of 3.5$\mu$eV, while the low capacitance tip has a bandwidth of 250$\mu$eV.

In summary, we have shown the benefit of using a high cross junction capacitance to perform JJS at elevated temperature by measuring on-chip Nb-based JJS devices at temperatures up to 4K. We also described how such large capacitances can be achieved in an STM, and characterized those tips experimentally and theoretically in the context of P(E)-theory.  These results provide a pathway for STM-JJS that can be used to characterize the nm-scale microwave absorption in samples and can be easily implemented on commercial STM systems without the need for \textit{in situ} microwave circuitry, complex impedance matching, or $\sim$mK temperatures. However, our results also demonstrate the need for low-pass filtering near the sample, which we showed was necessary to observe both the R-branch and the S-branch, in order for STM-JJS to be fully implemented.  Moreover, these experiments would benefit from a different material system, as the Dynes parameter is large for NbN possibly due to a thin passivation layer \cite{Brun, nbnlayer}. Higher capacitances could also be achieved by reducing the tip heights, which are limited by the roughness of the sample, or increasing the area of the tip.

\section*{Acknowledgments}

Funding was provided by Office of Naval Research award N00014-20-1-2356 (M.A.F., R.H.R. and V.W.B.). 
The work by M.A.F. was
also funded through a Google PhD Fellowship (Quantum Computing). 
This material is based upon research supported by the Chateaubriand Fellowship of the Office for Science \&
Technology of the Embassy of France in the United States (M.A.F.). Z.J.K. is supported by the U.S.
Department of Energy Office of Science National Quantum
Information Science Research Centers as part of the Q-NEXT
center. The authors would like to formally acknowledge and thank the Transatlantic Research Partnership, a program of the FACE Foundation and the French Embassy, for their support of this project (\c C.\"O.G.). Supported in part by the U.S. Department of Energy (DOE), Office of Science, Basic Energy Sciences (BES) under Award \#DE-SC0020313 (D.C.H. and R.M.). S.H. and M.S.J. acknowledge the support by the National Research Foundation of Korea (NRF) grant (RS-2024-00416583) funded by the Ministry of Science and ICT (MSIT) and the Korea Creative Content Agency grant (RS-2024-00332210) funded by the Ministry of Culture, Sports and Tourism (MCST).
The authors gratefully acknowledge the use of facilities and
instrumentation supported by NSF through the University of Wisconsin Materials Research Science and Engineering Center (No. DMR1720415).

\bibliography{biblio}

\begin{thebibliography}{61}%
\makeatletter
\providecommand \@ifxundefined [1]{%
 \@ifx{#1\undefined}
}%
\providecommand \@ifnum [1]{%
 \ifnum #1\expandafter \@firstoftwo
 \else \expandafter \@secondoftwo
 \fi
}%
\providecommand \@ifx [1]{%
 \ifx #1\expandafter \@firstoftwo
 \else \expandafter \@secondoftwo
 \fi
}%
\providecommand \natexlab [1]{#1}%
\providecommand \enquote  [1]{``#1''}%
\providecommand \bibnamefont  [1]{#1}%
\providecommand \bibfnamefont [1]{#1}%
\providecommand \citenamefont [1]{#1}%
\providecommand \href@noop [0]{\@secondoftwo}%
\providecommand \href [0]{\begingroup \@sanitize@url \@href}%
\providecommand \@href[1]{\@@startlink{#1}\@@href}%
\providecommand \@@href[1]{\endgroup#1\@@endlink}%
\providecommand \@sanitize@url [0]{\catcode `\\12\catcode `\$12\catcode `\&12\catcode `\#12\catcode `\^12\catcode `\_12\catcode `\%12\relax}%
\providecommand \@@startlink[1]{}%
\providecommand \@@endlink[0]{}%
\providecommand \url  [0]{\begingroup\@sanitize@url \@url }%
\providecommand \@url [1]{\endgroup\@href {#1}{\urlprefix }}%
\providecommand \urlprefix  [0]{URL }%
\providecommand \Eprint [0]{\href }%
\providecommand \doibase [0]{https://doi.org/}%
\providecommand \selectlanguage [0]{\@gobble}%
\providecommand \bibinfo  [0]{\@secondoftwo}%
\providecommand \bibfield  [0]{\@secondoftwo}%
\providecommand \translation [1]{[#1]}%
\providecommand \BibitemOpen [0]{}%
\providecommand \bibitemStop [0]{}%
\providecommand \bibitemNoStop [0]{.\EOS\space}%
\providecommand \EOS [0]{\spacefactor3000\relax}%
\providecommand \BibitemShut  [1]{\csname bibitem#1\endcsname}%
\let\auto@bib@innerbib\@empty
\bibitem [{\citenamefont {M{\"u}ller}\ \emph {et~al.}(2019)\citenamefont {M{\"u}ller}, \citenamefont {Cole},\ and\ \citenamefont {Lisenfeld}}]{Towardsunderstanding2019}%
  \BibitemOpen
  \bibfield  {author} {\bibinfo {author} {\bibfnamefont {C.}~\bibnamefont {M{\"u}ller}}, \bibinfo {author} {\bibfnamefont {J.~H.}\ \bibnamefont {Cole}},\ and\ \bibinfo {author} {\bibfnamefont {J.}~\bibnamefont {Lisenfeld}},\ }\bibfield  {title} {\bibinfo {title} {Towards understanding two-level-systems in amorphous solids: insights from quantum circuits},\ }\href {https://doi.org/10.1088/1361-6633/ab3a7e} {\bibfield  {journal} {\bibinfo  {journal} {Reports on Progress in Physics}\ }\textbf {\bibinfo {volume} {82}},\ \bibinfo {pages} {124501} (\bibinfo {year} {2019})}\BibitemShut {NoStop}%
\bibitem [{\citenamefont {Khalil}\ \emph {et~al.}(2014)\citenamefont {Khalil}, \citenamefont {Gladchenko}, \citenamefont {Stoutimore}, \citenamefont {Wellstood}, \citenamefont {Burin},\ and\ \citenamefont {Osborn}}]{Landau2014}%
  \BibitemOpen
  \bibfield  {author} {\bibinfo {author} {\bibfnamefont {M.~S.}\ \bibnamefont {Khalil}}, \bibinfo {author} {\bibfnamefont {S.}~\bibnamefont {Gladchenko}}, \bibinfo {author} {\bibfnamefont {M.~J.~A.}\ \bibnamefont {Stoutimore}}, \bibinfo {author} {\bibfnamefont {F.~C.}\ \bibnamefont {Wellstood}}, \bibinfo {author} {\bibfnamefont {A.~L.}\ \bibnamefont {Burin}},\ and\ \bibinfo {author} {\bibfnamefont {K.~D.}\ \bibnamefont {Osborn}},\ }\bibfield  {title} {\bibinfo {title} {Landau-zener population control and dipole measurement of a two-level-system bath},\ }\href {https://doi.org/10.1103/PhysRevB.90.100201} {\bibfield  {journal} {\bibinfo  {journal} {Phys. Rev. B}\ }\textbf {\bibinfo {volume} {90}},\ \bibinfo {pages} {100201} (\bibinfo {year} {2014})}\BibitemShut {NoStop}%
\bibitem [{\citenamefont {Sarabi}\ \emph {et~al.}(2016)\citenamefont {Sarabi}, \citenamefont {Ramanayaka}, \citenamefont {Burin}, \citenamefont {Wellstood},\ and\ \citenamefont {Osborn}}]{ProjectedDipole2016}%
  \BibitemOpen
  \bibfield  {author} {\bibinfo {author} {\bibfnamefont {B.}~\bibnamefont {Sarabi}}, \bibinfo {author} {\bibfnamefont {A.~N.}\ \bibnamefont {Ramanayaka}}, \bibinfo {author} {\bibfnamefont {A.~L.}\ \bibnamefont {Burin}}, \bibinfo {author} {\bibfnamefont {F.~C.}\ \bibnamefont {Wellstood}},\ and\ \bibinfo {author} {\bibfnamefont {K.~D.}\ \bibnamefont {Osborn}},\ }\bibfield  {title} {\bibinfo {title} {Projected dipole moments of individual two-level defects extracted using circuit quantum electrodynamics},\ }\href {https://doi.org/10.1103/PhysRevLett.116.167002} {\bibfield  {journal} {\bibinfo  {journal} {Phys. Rev. Lett.}\ }\textbf {\bibinfo {volume} {116}},\ \bibinfo {pages} {167002} (\bibinfo {year} {2016})}\BibitemShut {NoStop}%
\bibitem [{\citenamefont {Rosen}\ \emph {et~al.}(2016)\citenamefont {Rosen}, \citenamefont {Khalil}, \citenamefont {Burin},\ and\ \citenamefont {Osborn}}]{Laser2016}%
  \BibitemOpen
  \bibfield  {author} {\bibinfo {author} {\bibfnamefont {Y.~J.}\ \bibnamefont {Rosen}}, \bibinfo {author} {\bibfnamefont {M.~S.}\ \bibnamefont {Khalil}}, \bibinfo {author} {\bibfnamefont {A.~L.}\ \bibnamefont {Burin}},\ and\ \bibinfo {author} {\bibfnamefont {K.~D.}\ \bibnamefont {Osborn}},\ }\bibfield  {title} {\bibinfo {title} {Random-defect laser: Manipulating lossy two-level systems to produce a circuit with coherent gain},\ }\href {https://doi.org/10.1103/PhysRevLett.116.163601} {\bibfield  {journal} {\bibinfo  {journal} {Phys. Rev. Lett.}\ }\textbf {\bibinfo {volume} {116}},\ \bibinfo {pages} {163601} (\bibinfo {year} {2016})}\BibitemShut {NoStop}%
\bibitem [{\citenamefont {Brehm}\ \emph {et~al.}(2017)\citenamefont {Brehm}, \citenamefont {Bilmes}, \citenamefont {Weiss}, \citenamefont {Ustinov},\ and\ \citenamefont {Lisenfeld}}]{Transmissionline2017}%
  \BibitemOpen
  \bibfield  {author} {\bibinfo {author} {\bibfnamefont {J.~D.}\ \bibnamefont {Brehm}}, \bibinfo {author} {\bibfnamefont {A.}~\bibnamefont {Bilmes}}, \bibinfo {author} {\bibfnamefont {G.}~\bibnamefont {Weiss}}, \bibinfo {author} {\bibfnamefont {A.~V.}\ \bibnamefont {Ustinov}},\ and\ \bibinfo {author} {\bibfnamefont {J.}~\bibnamefont {Lisenfeld}},\ }\bibfield  {title} {\bibinfo {title} {Transmission-line resonators for the study of individual two-level tunneling systems},\ }\href {https://doi.org/10.1063/1.5001920} {\bibfield  {journal} {\bibinfo  {journal} {Applied Physics Letters}\ }\textbf {\bibinfo {volume} {111}},\ \bibinfo {pages} {112601} (\bibinfo {year} {2017})}\BibitemShut {NoStop}%
\bibitem [{\citenamefont {Matityahu}\ \emph {et~al.}(2019)\citenamefont {Matityahu}, \citenamefont {Schmidt}, \citenamefont {Bilmes}, \citenamefont {Shnirman}, \citenamefont {Weiss}, \citenamefont {Ustinov}, \citenamefont {Schechter},\ and\ \citenamefont {Lisenfeld}}]{Dynamical2019}%
  \BibitemOpen
  \bibfield  {author} {\bibinfo {author} {\bibfnamefont {S.}~\bibnamefont {Matityahu}}, \bibinfo {author} {\bibfnamefont {H.}~\bibnamefont {Schmidt}}, \bibinfo {author} {\bibfnamefont {A.}~\bibnamefont {Bilmes}}, \bibinfo {author} {\bibfnamefont {A.}~\bibnamefont {Shnirman}}, \bibinfo {author} {\bibfnamefont {G.}~\bibnamefont {Weiss}}, \bibinfo {author} {\bibfnamefont {A.~V.}\ \bibnamefont {Ustinov}}, \bibinfo {author} {\bibfnamefont {M.}~\bibnamefont {Schechter}},\ and\ \bibinfo {author} {\bibfnamefont {J.}~\bibnamefont {Lisenfeld}},\ }\bibfield  {title} {\bibinfo {title} {Dynamical decoupling of quantum two-level systems by coherent multiple landau--zener transitions},\ }\href {https://doi.org/10.1038/s41534-019-0228-x} {\bibfield  {journal} {\bibinfo  {journal} {npj Quantum Information}\ }\textbf {\bibinfo {volume} {5}},\ \bibinfo {pages} {114} (\bibinfo {year} {2019})}\BibitemShut {NoStop}%
\bibitem [{\citenamefont {Bilmes}\ \emph {et~al.}(2021)\citenamefont {Bilmes}, \citenamefont {Volosheniuk}, \citenamefont {Brehm}, \citenamefont {Ustinov},\ and\ \citenamefont {Lisenfeld}}]{Quantumsensors2021}%
  \BibitemOpen
  \bibfield  {author} {\bibinfo {author} {\bibfnamefont {A.}~\bibnamefont {Bilmes}}, \bibinfo {author} {\bibfnamefont {S.}~\bibnamefont {Volosheniuk}}, \bibinfo {author} {\bibfnamefont {J.~D.}\ \bibnamefont {Brehm}}, \bibinfo {author} {\bibfnamefont {A.~V.}\ \bibnamefont {Ustinov}},\ and\ \bibinfo {author} {\bibfnamefont {J.}~\bibnamefont {Lisenfeld}},\ }\bibfield  {title} {\bibinfo {title} {Quantum sensors for microscopic tunneling systems},\ }\href {https://doi.org/10.1038/s41534-020-00359-x} {\bibfield  {journal} {\bibinfo  {journal} {npj Quantum Information}\ }\textbf {\bibinfo {volume} {7}},\ \bibinfo {pages} {27} (\bibinfo {year} {2021})}\BibitemShut {NoStop}%
\bibitem [{\citenamefont {de~Graaf}\ \emph {et~al.}(2021)\citenamefont {de~Graaf}, \citenamefont {Mahashabde}, \citenamefont {Kubatkin}, \citenamefont {Tzalenchuk},\ and\ \citenamefont {Danilov}}]{Quantifying2021}%
  \BibitemOpen
  \bibfield  {author} {\bibinfo {author} {\bibfnamefont {S.~E.}\ \bibnamefont {de~Graaf}}, \bibinfo {author} {\bibfnamefont {S.}~\bibnamefont {Mahashabde}}, \bibinfo {author} {\bibfnamefont {S.~E.}\ \bibnamefont {Kubatkin}}, \bibinfo {author} {\bibfnamefont {A.~Y.}\ \bibnamefont {Tzalenchuk}},\ and\ \bibinfo {author} {\bibfnamefont {A.~V.}\ \bibnamefont {Danilov}},\ }\bibfield  {title} {\bibinfo {title} {Quantifying dynamics and interactions of individual spurious low-energy fluctuators in superconducting circuits},\ }\href {https://doi.org/10.1103/PhysRevB.103.174103} {\bibfield  {journal} {\bibinfo  {journal} {Phys. Rev. B}\ }\textbf {\bibinfo {volume} {103}},\ \bibinfo {pages} {174103} (\bibinfo {year} {2021})}\BibitemShut {NoStop}%
\bibitem [{\citenamefont {Yu}\ \emph {et~al.}(2022)\citenamefont {Yu}, \citenamefont {Matityahu}, \citenamefont {Rosen}, \citenamefont {Hung}, \citenamefont {Maksymov}, \citenamefont {Burin}, \citenamefont {Schechter},\ and\ \citenamefont {Osborn}}]{Experimentallyrevealing2022}%
  \BibitemOpen
  \bibfield  {author} {\bibinfo {author} {\bibfnamefont {L.}~\bibnamefont {Yu}}, \bibinfo {author} {\bibfnamefont {S.}~\bibnamefont {Matityahu}}, \bibinfo {author} {\bibfnamefont {Y.~J.}\ \bibnamefont {Rosen}}, \bibinfo {author} {\bibfnamefont {C.-C.}\ \bibnamefont {Hung}}, \bibinfo {author} {\bibfnamefont {A.}~\bibnamefont {Maksymov}}, \bibinfo {author} {\bibfnamefont {A.~L.}\ \bibnamefont {Burin}}, \bibinfo {author} {\bibfnamefont {M.}~\bibnamefont {Schechter}},\ and\ \bibinfo {author} {\bibfnamefont {K.~D.}\ \bibnamefont {Osborn}},\ }\bibfield  {title} {\bibinfo {title} {Experimentally revealing anomalously large dipoles in the dielectric of a quantum circuit},\ }\href {https://doi.org/10.1038/s41598-022-21256-7} {\bibfield  {journal} {\bibinfo  {journal} {Scientific Reports}\ }\textbf {\bibinfo {volume} {12}},\ \bibinfo {pages} {16960} (\bibinfo {year} {2022})}\BibitemShut {NoStop}%
\bibitem [{\citenamefont {Hung}\ \emph {et~al.}(2022)\citenamefont {Hung}, \citenamefont {Yu}, \citenamefont {Foroozani}, \citenamefont {Fritz}, \citenamefont {Gerthsen},\ and\ \citenamefont {Osborn}}]{Probinghundreds2022}%
  \BibitemOpen
  \bibfield  {author} {\bibinfo {author} {\bibfnamefont {C.-C.}\ \bibnamefont {Hung}}, \bibinfo {author} {\bibfnamefont {L.}~\bibnamefont {Yu}}, \bibinfo {author} {\bibfnamefont {N.}~\bibnamefont {Foroozani}}, \bibinfo {author} {\bibfnamefont {S.}~\bibnamefont {Fritz}}, \bibinfo {author} {\bibfnamefont {D.}~\bibnamefont {Gerthsen}},\ and\ \bibinfo {author} {\bibfnamefont {K.~D.}\ \bibnamefont {Osborn}},\ }\bibfield  {title} {\bibinfo {title} {Probing hundreds of individual quantum defects in polycrystalline and amorphous alumina},\ }\href {https://doi.org/10.1103/PhysRevApplied.17.034025} {\bibfield  {journal} {\bibinfo  {journal} {Phys. Rev. Appl.}\ }\textbf {\bibinfo {volume} {17}},\ \bibinfo {pages} {034025} (\bibinfo {year} {2022})}\BibitemShut {NoStop}%
\bibitem [{\citenamefont {Lisenfeld}\ \emph {et~al.}(2023)\citenamefont {Lisenfeld}, \citenamefont {Bilmes},\ and\ \citenamefont {Ustinov}}]{Enhancing2023}%
  \BibitemOpen
  \bibfield  {author} {\bibinfo {author} {\bibfnamefont {J.}~\bibnamefont {Lisenfeld}}, \bibinfo {author} {\bibfnamefont {A.}~\bibnamefont {Bilmes}},\ and\ \bibinfo {author} {\bibfnamefont {A.~V.}\ \bibnamefont {Ustinov}},\ }\bibfield  {title} {\bibinfo {title} {Enhancing the coherence of superconducting quantum bits with electric fields},\ }\href {https://doi.org/10.1038/s41534-023-00678-9} {\bibfield  {journal} {\bibinfo  {journal} {npj Quantum Information}\ }\textbf {\bibinfo {volume} {9}},\ \bibinfo {pages} {8} (\bibinfo {year} {2023})}\BibitemShut {NoStop}%
\bibitem [{\citenamefont {Burin}\ \emph {et~al.}(2024)\citenamefont {Burin}, \citenamefont {Schechter}, \citenamefont {Tennant}, \citenamefont {Ray},\ and\ \citenamefont {Rosen}}]{redshift2024}%
  \BibitemOpen
  \bibfield  {author} {\bibinfo {author} {\bibfnamefont {A.~L.}\ \bibnamefont {Burin}}, \bibinfo {author} {\bibfnamefont {M.}~\bibnamefont {Schechter}}, \bibinfo {author} {\bibfnamefont {D.}~\bibnamefont {Tennant}}, \bibinfo {author} {\bibfnamefont {K.~G.}\ \bibnamefont {Ray}},\ and\ \bibinfo {author} {\bibfnamefont {Y.~J.}\ \bibnamefont {Rosen}},\ }\href@noop {} {\bibinfo {title} {Red shift of the superconductivity cavity resonance in josephson junction qubits as a direct signature of tls population inversion}} (\bibinfo {year} {2024}),\ \Eprint {https://arxiv.org/abs/2401.15624} {arXiv:2401.15624 [cond-mat.dis-nn]} \BibitemShut {NoStop}%
\bibitem [{\citenamefont {Kristen}\ \emph {et~al.}(2024)\citenamefont {Kristen}, \citenamefont {Voss}, \citenamefont {Wildermuth}, \citenamefont {Bilmes}, \citenamefont {Lisenfeld}, \citenamefont {Rotzinger},\ and\ \citenamefont {Ustinov}}]{Giant2024}%
  \BibitemOpen
  \bibfield  {author} {\bibinfo {author} {\bibfnamefont {M.}~\bibnamefont {Kristen}}, \bibinfo {author} {\bibfnamefont {J.~N.}\ \bibnamefont {Voss}}, \bibinfo {author} {\bibfnamefont {M.}~\bibnamefont {Wildermuth}}, \bibinfo {author} {\bibfnamefont {A.}~\bibnamefont {Bilmes}}, \bibinfo {author} {\bibfnamefont {J.}~\bibnamefont {Lisenfeld}}, \bibinfo {author} {\bibfnamefont {H.}~\bibnamefont {Rotzinger}},\ and\ \bibinfo {author} {\bibfnamefont {A.~V.}\ \bibnamefont {Ustinov}},\ }\bibfield  {title} {\bibinfo {title} {Giant two-level systems in a granular superconductor},\ }\href {https://doi.org/10.1103/PhysRevLett.132.217002} {\bibfield  {journal} {\bibinfo  {journal} {Phys. Rev. Lett.}\ }\textbf {\bibinfo {volume} {132}},\ \bibinfo {pages} {217002} (\bibinfo {year} {2024})}\BibitemShut {NoStop}%
\bibitem [{\citenamefont {Grabovskij}\ \emph {et~al.}(2012)\citenamefont {Grabovskij}, \citenamefont {Peichl}, \citenamefont {Lisenfeld}, \citenamefont {Weiss},\ and\ \citenamefont {Ustinov}}]{StrainTuning2012}%
  \BibitemOpen
  \bibfield  {author} {\bibinfo {author} {\bibfnamefont {G.~J.}\ \bibnamefont {Grabovskij}}, \bibinfo {author} {\bibfnamefont {T.}~\bibnamefont {Peichl}}, \bibinfo {author} {\bibfnamefont {J.}~\bibnamefont {Lisenfeld}}, \bibinfo {author} {\bibfnamefont {G.}~\bibnamefont {Weiss}},\ and\ \bibinfo {author} {\bibfnamefont {A.~V.}\ \bibnamefont {Ustinov}},\ }\bibfield  {title} {\bibinfo {title} {Strain tuning of individual atomic tunneling systems detected by a superconducting qubit},\ }\href {https://doi.org/10.1126/science.1226487} {\bibfield  {journal} {\bibinfo  {journal} {Science}\ }\textbf {\bibinfo {volume} {338}},\ \bibinfo {pages} {232} (\bibinfo {year} {2012})}\BibitemShut {NoStop}%
\bibitem [{\citenamefont {Lisenfeld}\ \emph {et~al.}(2015)\citenamefont {Lisenfeld}, \citenamefont {Grabovskij}, \citenamefont {M{\"u}ller}, \citenamefont {Cole}, \citenamefont {Weiss},\ and\ \citenamefont {Ustinov}}]{Observation2015}%
  \BibitemOpen
  \bibfield  {author} {\bibinfo {author} {\bibfnamefont {J.}~\bibnamefont {Lisenfeld}}, \bibinfo {author} {\bibfnamefont {G.~J.}\ \bibnamefont {Grabovskij}}, \bibinfo {author} {\bibfnamefont {C.}~\bibnamefont {M{\"u}ller}}, \bibinfo {author} {\bibfnamefont {J.~H.}\ \bibnamefont {Cole}}, \bibinfo {author} {\bibfnamefont {G.}~\bibnamefont {Weiss}},\ and\ \bibinfo {author} {\bibfnamefont {A.~V.}\ \bibnamefont {Ustinov}},\ }\bibfield  {title} {\bibinfo {title} {Observation of directly interacting coherent two-level systems in an amorphous material},\ }\href {https://doi.org/10.1038/ncomms7182} {\bibfield  {journal} {\bibinfo  {journal} {Nature Communications}\ }\textbf {\bibinfo {volume} {6}},\ \bibinfo {pages} {6182} (\bibinfo {year} {2015})}\BibitemShut {NoStop}%
\bibitem [{\citenamefont {Bilmes}\ \emph {et~al.}(2022)\citenamefont {Bilmes}, \citenamefont {Volosheniuk}, \citenamefont {Ustinov},\ and\ \citenamefont {Lisenfeld}}]{Probingdefect2022}%
  \BibitemOpen
  \bibfield  {author} {\bibinfo {author} {\bibfnamefont {A.}~\bibnamefont {Bilmes}}, \bibinfo {author} {\bibfnamefont {S.}~\bibnamefont {Volosheniuk}}, \bibinfo {author} {\bibfnamefont {A.~V.}\ \bibnamefont {Ustinov}},\ and\ \bibinfo {author} {\bibfnamefont {J.}~\bibnamefont {Lisenfeld}},\ }\bibfield  {title} {\bibinfo {title} {Probing defect densities at the edges and inside josephson junctions of superconducting qubits},\ }\href {https://doi.org/10.1038/s41534-022-00532-4} {\bibfield  {journal} {\bibinfo  {journal} {npj Quantum Information}\ }\textbf {\bibinfo {volume} {8}},\ \bibinfo {pages} {24} (\bibinfo {year} {2022})}\BibitemShut {NoStop}%
\bibitem [{\citenamefont {Wenner}\ \emph {et~al.}(2011)\citenamefont {Wenner}, \citenamefont {Barends}, \citenamefont {Bialczak}, \citenamefont {Chen}, \citenamefont {Kelly}, \citenamefont {Lucero}, \citenamefont {Mariantoni}, \citenamefont {Megrant}, \citenamefont {O'Malley}, \citenamefont {Sank}, \citenamefont {Vainsencher}, \citenamefont {Wang}, \citenamefont {White}, \citenamefont {Yin}, \citenamefont {Zhao}, \citenamefont {Cleland},\ and\ \citenamefont {Martinis}}]{Wenner2011}%
  \BibitemOpen
  \bibfield  {author} {\bibinfo {author} {\bibfnamefont {J.}~\bibnamefont {Wenner}}, \bibinfo {author} {\bibfnamefont {R.}~\bibnamefont {Barends}}, \bibinfo {author} {\bibfnamefont {R.~C.}\ \bibnamefont {Bialczak}}, \bibinfo {author} {\bibfnamefont {Y.}~\bibnamefont {Chen}}, \bibinfo {author} {\bibfnamefont {J.}~\bibnamefont {Kelly}}, \bibinfo {author} {\bibfnamefont {E.}~\bibnamefont {Lucero}}, \bibinfo {author} {\bibfnamefont {M.}~\bibnamefont {Mariantoni}}, \bibinfo {author} {\bibfnamefont {A.}~\bibnamefont {Megrant}}, \bibinfo {author} {\bibfnamefont {P.~J.~J.}\ \bibnamefont {O'Malley}}, \bibinfo {author} {\bibfnamefont {D.}~\bibnamefont {Sank}}, \bibinfo {author} {\bibfnamefont {A.}~\bibnamefont {Vainsencher}}, \bibinfo {author} {\bibfnamefont {H.}~\bibnamefont {Wang}}, \bibinfo {author} {\bibfnamefont {T.~C.}\ \bibnamefont {White}}, \bibinfo {author} {\bibfnamefont {Y.}~\bibnamefont {Yin}}, \bibinfo {author} {\bibfnamefont {J.}~\bibnamefont {Zhao}}, \bibinfo {author} {\bibfnamefont {A.~N.}\ \bibnamefont
  {Cleland}},\ and\ \bibinfo {author} {\bibfnamefont {J.~M.}\ \bibnamefont {Martinis}},\ }\bibfield  {title} {\bibinfo {title} {Surface loss simulations of superconducting coplanar waveguide resonators},\ }\href {https://doi.org/10.1063/1.3637047} {\bibfield  {journal} {\bibinfo  {journal} {Applied Physics Letters}\ }\textbf {\bibinfo {volume} {99}},\ \bibinfo {pages} {113513} (\bibinfo {year} {2011})}\BibitemShut {NoStop}%
\bibitem [{\citenamefont {Sandberg}\ \emph {et~al.}(2013)\citenamefont {Sandberg}, \citenamefont {Vissers}, \citenamefont {Ohki}, \citenamefont {Gao}, \citenamefont {Aumentado}, \citenamefont {Weides},\ and\ \citenamefont {Pappas}}]{Sandberg2013}%
  \BibitemOpen
  \bibfield  {author} {\bibinfo {author} {\bibfnamefont {M.}~\bibnamefont {Sandberg}}, \bibinfo {author} {\bibfnamefont {M.~R.}\ \bibnamefont {Vissers}}, \bibinfo {author} {\bibfnamefont {T.~A.}\ \bibnamefont {Ohki}}, \bibinfo {author} {\bibfnamefont {J.}~\bibnamefont {Gao}}, \bibinfo {author} {\bibfnamefont {J.}~\bibnamefont {Aumentado}}, \bibinfo {author} {\bibfnamefont {M.}~\bibnamefont {Weides}},\ and\ \bibinfo {author} {\bibfnamefont {D.~P.}\ \bibnamefont {Pappas}},\ }\bibfield  {title} {\bibinfo {title} {Radiation-suppressed superconducting quantum bit in a planar geometry},\ }\href {https://doi.org/10.1063/1.4792698} {\bibfield  {journal} {\bibinfo  {journal} {Applied Physics Letters}\ }\textbf {\bibinfo {volume} {102}},\ \bibinfo {pages} {072601} (\bibinfo {year} {2013})}\BibitemShut {NoStop}%
\bibitem [{\citenamefont {Wang}\ \emph {et~al.}(2015)\citenamefont {Wang}, \citenamefont {Axline}, \citenamefont {Gao}, \citenamefont {Brecht}, \citenamefont {Chu}, \citenamefont {Frunzio}, \citenamefont {Devoret},\ and\ \citenamefont {Schoelkopf}}]{surfacepart2015}%
  \BibitemOpen
  \bibfield  {author} {\bibinfo {author} {\bibfnamefont {C.}~\bibnamefont {Wang}}, \bibinfo {author} {\bibfnamefont {C.}~\bibnamefont {Axline}}, \bibinfo {author} {\bibfnamefont {Y.~Y.}\ \bibnamefont {Gao}}, \bibinfo {author} {\bibfnamefont {T.}~\bibnamefont {Brecht}}, \bibinfo {author} {\bibfnamefont {Y.}~\bibnamefont {Chu}}, \bibinfo {author} {\bibfnamefont {L.}~\bibnamefont {Frunzio}}, \bibinfo {author} {\bibfnamefont {M.~H.}\ \bibnamefont {Devoret}},\ and\ \bibinfo {author} {\bibfnamefont {R.~J.}\ \bibnamefont {Schoelkopf}},\ }\bibfield  {title} {\bibinfo {title} {Surface participation and dielectric loss in superconducting qubits},\ }\href {https://doi.org/10.1063/1.4934486} {\bibfield  {journal} {\bibinfo  {journal} {Applied Physics Letters}\ }\textbf {\bibinfo {volume} {107}},\ \bibinfo {pages} {162601} (\bibinfo {year} {2015})}\BibitemShut {NoStop}%
\bibitem [{\citenamefont {Dial}\ \emph {et~al.}(2016{\natexlab{a}})\citenamefont {Dial}, \citenamefont {McClure}, \citenamefont {Poletto}, \citenamefont {Keefe}, \citenamefont {Rothwell}, \citenamefont {Gambetta}, \citenamefont {Abraham}, \citenamefont {Chow},\ and\ \citenamefont {Steffen}}]{Bulkandsurfaceloss2016}%
  \BibitemOpen
  \bibfield  {author} {\bibinfo {author} {\bibfnamefont {O.}~\bibnamefont {Dial}}, \bibinfo {author} {\bibfnamefont {D.~T.}\ \bibnamefont {McClure}}, \bibinfo {author} {\bibfnamefont {S.}~\bibnamefont {Poletto}}, \bibinfo {author} {\bibfnamefont {G.~A.}\ \bibnamefont {Keefe}}, \bibinfo {author} {\bibfnamefont {M.~B.}\ \bibnamefont {Rothwell}}, \bibinfo {author} {\bibfnamefont {J.~M.}\ \bibnamefont {Gambetta}}, \bibinfo {author} {\bibfnamefont {D.~W.}\ \bibnamefont {Abraham}}, \bibinfo {author} {\bibfnamefont {J.~M.}\ \bibnamefont {Chow}},\ and\ \bibinfo {author} {\bibfnamefont {M.}~\bibnamefont {Steffen}},\ }\bibfield  {title} {\bibinfo {title} {Bulk and surface loss in superconducting transmon qubits},\ }\href {https://doi.org/10.1088/0953-2048/29/4/044001} {\bibfield  {journal} {\bibinfo  {journal} {Superconductor Science and Technology}\ }\textbf {\bibinfo {volume} {29}},\ \bibinfo {pages} {044001} (\bibinfo {year} {2016}{\natexlab{a}})}\BibitemShut {NoStop}%
\bibitem [{\citenamefont {Dial}\ \emph {et~al.}(2016{\natexlab{b}})\citenamefont {Dial}, \citenamefont {McClure}, \citenamefont {Poletto}, \citenamefont {Keefe}, \citenamefont {Rothwell}, \citenamefont {Gambetta}, \citenamefont {Abraham}, \citenamefont {Chow},\ and\ \citenamefont {Steffen}}]{Dial2016}%
  \BibitemOpen
  \bibfield  {author} {\bibinfo {author} {\bibfnamefont {O.}~\bibnamefont {Dial}}, \bibinfo {author} {\bibfnamefont {D.~T.}\ \bibnamefont {McClure}}, \bibinfo {author} {\bibfnamefont {S.}~\bibnamefont {Poletto}}, \bibinfo {author} {\bibfnamefont {G.~A.}\ \bibnamefont {Keefe}}, \bibinfo {author} {\bibfnamefont {M.~B.}\ \bibnamefont {Rothwell}}, \bibinfo {author} {\bibfnamefont {J.~M.}\ \bibnamefont {Gambetta}}, \bibinfo {author} {\bibfnamefont {D.~W.}\ \bibnamefont {Abraham}}, \bibinfo {author} {\bibfnamefont {J.~M.}\ \bibnamefont {Chow}},\ and\ \bibinfo {author} {\bibfnamefont {M.}~\bibnamefont {Steffen}},\ }\bibfield  {title} {\bibinfo {title} {Bulk and surface loss in superconducting transmon qubits},\ }\href {https://doi.org/10.1088/0953-2048/29/4/044001} {\bibfield  {journal} {\bibinfo  {journal} {Superconductor Science and Technology}\ }\textbf {\bibinfo {volume} {29}},\ \bibinfo {pages} {044001} (\bibinfo {year} {2016}{\natexlab{b}})}\BibitemShut {NoStop}%
\bibitem [{\citenamefont {Gambetta}\ \emph {et~al.}(2017)\citenamefont {Gambetta}, \citenamefont {Murray}, \citenamefont {.~K.~Fung}, \citenamefont {McClure}, \citenamefont {Dial}, \citenamefont {Shanks}, \citenamefont {Sleight},\ and\ \citenamefont {Steffen}}]{InvestigatingSurface2017}%
  \BibitemOpen
  \bibfield  {author} {\bibinfo {author} {\bibfnamefont {J.~M.}\ \bibnamefont {Gambetta}}, \bibinfo {author} {\bibfnamefont {C.~E.}\ \bibnamefont {Murray}}, \bibinfo {author} {\bibfnamefont {Y.-K.}\ \bibnamefont {.~K.~Fung}}, \bibinfo {author} {\bibfnamefont {D.~T.}\ \bibnamefont {McClure}}, \bibinfo {author} {\bibfnamefont {O.}~\bibnamefont {Dial}}, \bibinfo {author} {\bibfnamefont {W.}~\bibnamefont {Shanks}}, \bibinfo {author} {\bibfnamefont {J.~W.}\ \bibnamefont {Sleight}},\ and\ \bibinfo {author} {\bibfnamefont {M.}~\bibnamefont {Steffen}},\ }\bibfield  {title} {\bibinfo {title} {Investigating surface loss effects in superconducting transmon qubits},\ }\href {https://doi.org/10.1109/TASC.2016.2629670} {\bibfield  {journal} {\bibinfo  {journal} {IEEE Transactions on Applied Superconductivity}\ }\textbf {\bibinfo {volume} {27}},\ \bibinfo {pages} {1} (\bibinfo {year} {2017})}\BibitemShut {NoStop}%
\bibitem [{\citenamefont {Calusine}\ \emph {et~al.}(2018)\citenamefont {Calusine}, \citenamefont {Melville}, \citenamefont {Woods}, \citenamefont {Das}, \citenamefont {Stull}, \citenamefont {Bolkhovsky}, \citenamefont {Braje}, \citenamefont {Hover}, \citenamefont {Kim}, \citenamefont {Miloshi}, \citenamefont {Rosenberg}, \citenamefont {Sevi}, \citenamefont {Yoder}, \citenamefont {Dauler},\ and\ \citenamefont {Oliver}}]{AnalysisandMitigation2018}%
  \BibitemOpen
  \bibfield  {author} {\bibinfo {author} {\bibfnamefont {G.}~\bibnamefont {Calusine}}, \bibinfo {author} {\bibfnamefont {A.}~\bibnamefont {Melville}}, \bibinfo {author} {\bibfnamefont {W.}~\bibnamefont {Woods}}, \bibinfo {author} {\bibfnamefont {R.}~\bibnamefont {Das}}, \bibinfo {author} {\bibfnamefont {C.}~\bibnamefont {Stull}}, \bibinfo {author} {\bibfnamefont {V.}~\bibnamefont {Bolkhovsky}}, \bibinfo {author} {\bibfnamefont {D.}~\bibnamefont {Braje}}, \bibinfo {author} {\bibfnamefont {D.}~\bibnamefont {Hover}}, \bibinfo {author} {\bibfnamefont {D.~K.}\ \bibnamefont {Kim}}, \bibinfo {author} {\bibfnamefont {X.}~\bibnamefont {Miloshi}}, \bibinfo {author} {\bibfnamefont {D.}~\bibnamefont {Rosenberg}}, \bibinfo {author} {\bibfnamefont {A.}~\bibnamefont {Sevi}}, \bibinfo {author} {\bibfnamefont {J.~L.}\ \bibnamefont {Yoder}}, \bibinfo {author} {\bibfnamefont {E.}~\bibnamefont {Dauler}},\ and\ \bibinfo {author} {\bibfnamefont {W.~D.}\ \bibnamefont {Oliver}},\ }\bibfield  {title} {\bibinfo {title} {Analysis and
  mitigation of interface losses in trenched superconducting coplanar waveguide resonators},\ }\href {https://doi.org/10.1063/1.5006888} {\bibfield  {journal} {\bibinfo  {journal} {Applied Physics Letters}\ }\textbf {\bibinfo {volume} {112}},\ \bibinfo {pages} {062601} (\bibinfo {year} {2018})}\BibitemShut {NoStop}%
\bibitem [{\citenamefont {Woods}\ \emph {et~al.}(2019)\citenamefont {Woods}, \citenamefont {Calusine}, \citenamefont {Melville}, \citenamefont {Sevi}, \citenamefont {Golden}, \citenamefont {Kim}, \citenamefont {Rosenberg}, \citenamefont {Yoder},\ and\ \citenamefont {Oliver}}]{DeterminingInterface2019}%
  \BibitemOpen
  \bibfield  {author} {\bibinfo {author} {\bibfnamefont {W.}~\bibnamefont {Woods}}, \bibinfo {author} {\bibfnamefont {G.}~\bibnamefont {Calusine}}, \bibinfo {author} {\bibfnamefont {A.}~\bibnamefont {Melville}}, \bibinfo {author} {\bibfnamefont {A.}~\bibnamefont {Sevi}}, \bibinfo {author} {\bibfnamefont {E.}~\bibnamefont {Golden}}, \bibinfo {author} {\bibfnamefont {D.}~\bibnamefont {Kim}}, \bibinfo {author} {\bibfnamefont {D.}~\bibnamefont {Rosenberg}}, \bibinfo {author} {\bibfnamefont {J.}~\bibnamefont {Yoder}},\ and\ \bibinfo {author} {\bibfnamefont {W.}~\bibnamefont {Oliver}},\ }\bibfield  {title} {\bibinfo {title} {Determining interface dielectric losses in superconducting coplanar-waveguide resonators},\ }\href {https://doi.org/10.1103/PhysRevApplied.12.014012} {\bibfield  {journal} {\bibinfo  {journal} {Phys. Rev. Appl.}\ }\textbf {\bibinfo {volume} {12}},\ \bibinfo {pages} {014012} (\bibinfo {year} {2019})}\BibitemShut {NoStop}%
\bibitem [{\citenamefont {Verjauw}\ \emph {et~al.}(2022)\citenamefont {Verjauw}, \citenamefont {Acharya}, \citenamefont {Van~Damme}, \citenamefont {Ivanov}, \citenamefont {Lozano}, \citenamefont {Mohiyaddin}, \citenamefont {Wan}, \citenamefont {Jussot}, \citenamefont {Vadiraj}, \citenamefont {Mongillo}, \citenamefont {Heyns}, \citenamefont {Radu}, \citenamefont {Govoreanu},\ and\ \citenamefont {Poto{\v{c}}nik}}]{Verjauw2022}%
  \BibitemOpen
  \bibfield  {author} {\bibinfo {author} {\bibfnamefont {J.}~\bibnamefont {Verjauw}}, \bibinfo {author} {\bibfnamefont {R.}~\bibnamefont {Acharya}}, \bibinfo {author} {\bibfnamefont {J.}~\bibnamefont {Van~Damme}}, \bibinfo {author} {\bibfnamefont {T.}~\bibnamefont {Ivanov}}, \bibinfo {author} {\bibfnamefont {D.~P.}\ \bibnamefont {Lozano}}, \bibinfo {author} {\bibfnamefont {F.~A.}\ \bibnamefont {Mohiyaddin}}, \bibinfo {author} {\bibfnamefont {D.}~\bibnamefont {Wan}}, \bibinfo {author} {\bibfnamefont {J.}~\bibnamefont {Jussot}}, \bibinfo {author} {\bibfnamefont {A.~M.}\ \bibnamefont {Vadiraj}}, \bibinfo {author} {\bibfnamefont {M.}~\bibnamefont {Mongillo}}, \bibinfo {author} {\bibfnamefont {M.}~\bibnamefont {Heyns}}, \bibinfo {author} {\bibfnamefont {I.}~\bibnamefont {Radu}}, \bibinfo {author} {\bibfnamefont {B.}~\bibnamefont {Govoreanu}},\ and\ \bibinfo {author} {\bibfnamefont {A.}~\bibnamefont {Poto{\v{c}}nik}},\ }\bibfield  {title} {\bibinfo {title} {Path toward manufacturable superconducting qubits with
  relaxation times exceeding 0.1{\thinspace}ms},\ }\href {https://doi.org/10.1038/s41534-022-00600-9} {\bibfield  {journal} {\bibinfo  {journal} {npj Quantum Information}\ }\textbf {\bibinfo {volume} {8}},\ \bibinfo {pages} {93} (\bibinfo {year} {2022})}\BibitemShut {NoStop}%
\bibitem [{\citenamefont {Wang}\ \emph {et~al.}(2022)\citenamefont {Wang}, \citenamefont {Li}, \citenamefont {Xu}, \citenamefont {Li}, \citenamefont {Wang}, \citenamefont {Yang}, \citenamefont {Mi}, \citenamefont {Liang}, \citenamefont {Su}, \citenamefont {Yang}, \citenamefont {Wang}, \citenamefont {Wang}, \citenamefont {Li}, \citenamefont {Chen}, \citenamefont {Li}, \citenamefont {Linghu}, \citenamefont {Han}, \citenamefont {Zhang}, \citenamefont {Feng}, \citenamefont {Song}, \citenamefont {Ma}, \citenamefont {Zhang}, \citenamefont {Wang}, \citenamefont {Zhao}, \citenamefont {Liu}, \citenamefont {Xue}, \citenamefont {Jin},\ and\ \citenamefont {Yu}}]{Wang2022}%
  \BibitemOpen
  \bibfield  {author} {\bibinfo {author} {\bibfnamefont {C.}~\bibnamefont {Wang}}, \bibinfo {author} {\bibfnamefont {X.}~\bibnamefont {Li}}, \bibinfo {author} {\bibfnamefont {H.}~\bibnamefont {Xu}}, \bibinfo {author} {\bibfnamefont {Z.}~\bibnamefont {Li}}, \bibinfo {author} {\bibfnamefont {J.}~\bibnamefont {Wang}}, \bibinfo {author} {\bibfnamefont {Z.}~\bibnamefont {Yang}}, \bibinfo {author} {\bibfnamefont {Z.}~\bibnamefont {Mi}}, \bibinfo {author} {\bibfnamefont {X.}~\bibnamefont {Liang}}, \bibinfo {author} {\bibfnamefont {T.}~\bibnamefont {Su}}, \bibinfo {author} {\bibfnamefont {C.}~\bibnamefont {Yang}}, \bibinfo {author} {\bibfnamefont {G.}~\bibnamefont {Wang}}, \bibinfo {author} {\bibfnamefont {W.}~\bibnamefont {Wang}}, \bibinfo {author} {\bibfnamefont {Y.}~\bibnamefont {Li}}, \bibinfo {author} {\bibfnamefont {M.}~\bibnamefont {Chen}}, \bibinfo {author} {\bibfnamefont {C.}~\bibnamefont {Li}}, \bibinfo {author} {\bibfnamefont {K.}~\bibnamefont {Linghu}}, \bibinfo {author} {\bibfnamefont {J.}~\bibnamefont
  {Han}}, \bibinfo {author} {\bibfnamefont {Y.}~\bibnamefont {Zhang}}, \bibinfo {author} {\bibfnamefont {Y.}~\bibnamefont {Feng}}, \bibinfo {author} {\bibfnamefont {Y.}~\bibnamefont {Song}}, \bibinfo {author} {\bibfnamefont {T.}~\bibnamefont {Ma}}, \bibinfo {author} {\bibfnamefont {J.}~\bibnamefont {Zhang}}, \bibinfo {author} {\bibfnamefont {R.}~\bibnamefont {Wang}}, \bibinfo {author} {\bibfnamefont {P.}~\bibnamefont {Zhao}}, \bibinfo {author} {\bibfnamefont {W.}~\bibnamefont {Liu}}, \bibinfo {author} {\bibfnamefont {G.}~\bibnamefont {Xue}}, \bibinfo {author} {\bibfnamefont {Y.}~\bibnamefont {Jin}},\ and\ \bibinfo {author} {\bibfnamefont {H.}~\bibnamefont {Yu}},\ }\bibfield  {title} {\bibinfo {title} {Towards practical quantum computers: transmon qubit with a lifetime approaching 0.5 milliseconds},\ }\href {https://doi.org/10.1038/s41534-021-00510-2} {\bibfield  {journal} {\bibinfo  {journal} {npj Quantum Information}\ }\textbf {\bibinfo {volume} {8}},\ \bibinfo {pages} {3} (\bibinfo {year}
  {2022})}\BibitemShut {NoStop}%
\bibitem [{\citenamefont {Eun}\ \emph {et~al.}(2023)\citenamefont {Eun}, \citenamefont {Park}, \citenamefont {Seo}, \citenamefont {Choi},\ and\ \citenamefont {Hahn}}]{Eun2023}%
  \BibitemOpen
  \bibfield  {author} {\bibinfo {author} {\bibfnamefont {S.}~\bibnamefont {Eun}}, \bibinfo {author} {\bibfnamefont {S.~H.}\ \bibnamefont {Park}}, \bibinfo {author} {\bibfnamefont {K.}~\bibnamefont {Seo}}, \bibinfo {author} {\bibfnamefont {K.}~\bibnamefont {Choi}},\ and\ \bibinfo {author} {\bibfnamefont {S.}~\bibnamefont {Hahn}},\ }\bibfield  {title} {\bibinfo {title} {Shape optimization of superconducting transmon qubits for low surface dielectric loss},\ }\href {https://doi.org/10.1088/1361-6463/acf7cf} {\bibfield  {journal} {\bibinfo  {journal} {Journal of Physics D: Applied Physics}\ }\textbf {\bibinfo {volume} {56}},\ \bibinfo {pages} {505306} (\bibinfo {year} {2023})}\BibitemShut {NoStop}%
\bibitem [{\citenamefont {Lisenfeld}\ \emph {et~al.}(2019)\citenamefont {Lisenfeld}, \citenamefont {Bilmes}, \citenamefont {Megrant}, \citenamefont {Barends}, \citenamefont {Kelly}, \citenamefont {Klimov}, \citenamefont {Weiss}, \citenamefont {Martinis},\ and\ \citenamefont {Ustinov}}]{efieldspec2019}%
  \BibitemOpen
  \bibfield  {author} {\bibinfo {author} {\bibfnamefont {J.}~\bibnamefont {Lisenfeld}}, \bibinfo {author} {\bibfnamefont {A.}~\bibnamefont {Bilmes}}, \bibinfo {author} {\bibfnamefont {A.}~\bibnamefont {Megrant}}, \bibinfo {author} {\bibfnamefont {R.}~\bibnamefont {Barends}}, \bibinfo {author} {\bibfnamefont {J.}~\bibnamefont {Kelly}}, \bibinfo {author} {\bibfnamefont {P.}~\bibnamefont {Klimov}}, \bibinfo {author} {\bibfnamefont {G.}~\bibnamefont {Weiss}}, \bibinfo {author} {\bibfnamefont {J.~M.}\ \bibnamefont {Martinis}},\ and\ \bibinfo {author} {\bibfnamefont {A.~V.}\ \bibnamefont {Ustinov}},\ }\bibfield  {title} {\bibinfo {title} {Electric field spectroscopy of material defects in transmon qubits},\ }\href {https://doi.org/10.1038/s41534-019-0224-1} {\bibfield  {journal} {\bibinfo  {journal} {npj Quantum Information}\ }\textbf {\bibinfo {volume} {5}},\ \bibinfo {pages} {105} (\bibinfo {year} {2019})}\BibitemShut {NoStop}%
\bibitem [{\citenamefont {Bilmes}\ \emph {et~al.}(2020)\citenamefont {Bilmes}, \citenamefont {Megrant}, \citenamefont {Klimov}, \citenamefont {Weiss}, \citenamefont {Martinis}, \citenamefont {Ustinov},\ and\ \citenamefont {Lisenfeld}}]{Resolving2020}%
  \BibitemOpen
  \bibfield  {author} {\bibinfo {author} {\bibfnamefont {A.}~\bibnamefont {Bilmes}}, \bibinfo {author} {\bibfnamefont {A.}~\bibnamefont {Megrant}}, \bibinfo {author} {\bibfnamefont {P.}~\bibnamefont {Klimov}}, \bibinfo {author} {\bibfnamefont {G.}~\bibnamefont {Weiss}}, \bibinfo {author} {\bibfnamefont {J.~M.}\ \bibnamefont {Martinis}}, \bibinfo {author} {\bibfnamefont {A.~V.}\ \bibnamefont {Ustinov}},\ and\ \bibinfo {author} {\bibfnamefont {J.}~\bibnamefont {Lisenfeld}},\ }\bibfield  {title} {\bibinfo {title} {Resolving the positions of defects in superconducting quantum bits},\ }\href {https://doi.org/10.1038/s41598-020-59749-y} {\bibfield  {journal} {\bibinfo  {journal} {Scientific Reports}\ }\textbf {\bibinfo {volume} {10}},\ \bibinfo {pages} {3090} (\bibinfo {year} {2020})}\BibitemShut {NoStop}%
\bibitem [{\citenamefont {Bombin}(2010)}]{majorana1}%
  \BibitemOpen
  \bibfield  {author} {\bibinfo {author} {\bibfnamefont {H.}~\bibnamefont {Bombin}},\ }\bibfield  {title} {\bibinfo {title} {Topological order with a twist: Ising anyons from an abelian model},\ }\href {https://doi.org/10.1103/PhysRevLett.105.030403} {\bibfield  {journal} {\bibinfo  {journal} {Phys. Rev. Lett.}\ }\textbf {\bibinfo {volume} {105}},\ \bibinfo {pages} {030403} (\bibinfo {year} {2010})}\BibitemShut {NoStop}%
\bibitem [{\citenamefont {Sau}\ \emph {et~al.}(2015)\citenamefont {Sau}, \citenamefont {Swingle},\ and\ \citenamefont {Tewari}}]{majorana2}%
  \BibitemOpen
  \bibfield  {author} {\bibinfo {author} {\bibfnamefont {J.~D.}\ \bibnamefont {Sau}}, \bibinfo {author} {\bibfnamefont {B.}~\bibnamefont {Swingle}},\ and\ \bibinfo {author} {\bibfnamefont {S.}~\bibnamefont {Tewari}},\ }\bibfield  {title} {\bibinfo {title} {Proposal to probe quantum nonlocality of majorana fermions in tunneling experiments},\ }\href {https://doi.org/10.1103/PhysRevB.92.020511} {\bibfield  {journal} {\bibinfo  {journal} {Phys. Rev. B}\ }\textbf {\bibinfo {volume} {92}},\ \bibinfo {pages} {020511} (\bibinfo {year} {2015})}\BibitemShut {NoStop}%
\bibitem [{\citenamefont {Houzet}\ \emph {et~al.}(2013)\citenamefont {Houzet}, \citenamefont {Meyer}, \citenamefont {Badiane},\ and\ \citenamefont {Glazman}}]{majorana3}%
  \BibitemOpen
  \bibfield  {author} {\bibinfo {author} {\bibfnamefont {M.}~\bibnamefont {Houzet}}, \bibinfo {author} {\bibfnamefont {J.~S.}\ \bibnamefont {Meyer}}, \bibinfo {author} {\bibfnamefont {D.~M.}\ \bibnamefont {Badiane}},\ and\ \bibinfo {author} {\bibfnamefont {L.~I.}\ \bibnamefont {Glazman}},\ }\bibfield  {title} {\bibinfo {title} {Dynamics of majorana states in a topological josephson junction},\ }\href {https://doi.org/10.1103/PhysRevLett.111.046401} {\bibfield  {journal} {\bibinfo  {journal} {Phys. Rev. Lett.}\ }\textbf {\bibinfo {volume} {111}},\ \bibinfo {pages} {046401} (\bibinfo {year} {2013})}\BibitemShut {NoStop}%
\bibitem [{\citenamefont {Virtanen}\ and\ \citenamefont {Recher}(2013)}]{majorana4}%
  \BibitemOpen
  \bibfield  {author} {\bibinfo {author} {\bibfnamefont {P.}~\bibnamefont {Virtanen}}\ and\ \bibinfo {author} {\bibfnamefont {P.}~\bibnamefont {Recher}},\ }\bibfield  {title} {\bibinfo {title} {Microwave spectroscopy of josephson junctions in topological superconductors},\ }\href {https://doi.org/10.1103/PhysRevB.88.144507} {\bibfield  {journal} {\bibinfo  {journal} {Phys. Rev. B}\ }\textbf {\bibinfo {volume} {88}},\ \bibinfo {pages} {144507} (\bibinfo {year} {2013})}\BibitemShut {NoStop}%
\bibitem [{\citenamefont {Zhang}(2018)}]{majorana5}%
  \BibitemOpen
  \bibfield  {author} {\bibinfo {author} {\bibfnamefont {Z.-T.}\ \bibnamefont {Zhang}},\ }\bibfield  {title} {\bibinfo {title} {Distinguishing majorana bound states and andreev bound states with microwave spectra},\ }\href {https://doi.org/10.1088/1361-648X/aab134} {\bibfield  {journal} {\bibinfo  {journal} {Journal of Physics: Condensed Matter}\ }\textbf {\bibinfo {volume} {30}},\ \bibinfo {pages} {145402} (\bibinfo {year} {2018})}\BibitemShut {NoStop}%
\bibitem [{\citenamefont {Kemiktarak}\ \emph {et~al.}(2007)\citenamefont {Kemiktarak}, \citenamefont {Ndukum}, \citenamefont {Schwab},\ and\ \citenamefont {Ekinci}}]{RFSTM2007}%
  \BibitemOpen
  \bibfield  {author} {\bibinfo {author} {\bibfnamefont {U.}~\bibnamefont {Kemiktarak}}, \bibinfo {author} {\bibfnamefont {T.}~\bibnamefont {Ndukum}}, \bibinfo {author} {\bibfnamefont {K.~C.}\ \bibnamefont {Schwab}},\ and\ \bibinfo {author} {\bibfnamefont {K.~L.}\ \bibnamefont {Ekinci}},\ }\bibfield  {title} {\bibinfo {title} {Radio-frequency scanning tunnelling microscopy},\ }\href {https://doi.org/10.1038/nature06238} {\bibfield  {journal} {\bibinfo  {journal} {Nature}\ }\textbf {\bibinfo {volume} {450}},\ \bibinfo {pages} {85} (\bibinfo {year} {2007})}\BibitemShut {NoStop}%
\bibitem [{\citenamefont {Cao}\ \emph {et~al.}(2023)\citenamefont {Cao}, \citenamefont {Wu}, \citenamefont {Bhattacharyya}, \citenamefont {Zhang},\ and\ \citenamefont {Allen}}]{MAllen2023}%
  \BibitemOpen
  \bibfield  {author} {\bibinfo {author} {\bibfnamefont {L.~W.}\ \bibnamefont {Cao}}, \bibinfo {author} {\bibfnamefont {C.}~\bibnamefont {Wu}}, \bibinfo {author} {\bibfnamefont {R.}~\bibnamefont {Bhattacharyya}}, \bibinfo {author} {\bibfnamefont {R.}~\bibnamefont {Zhang}},\ and\ \bibinfo {author} {\bibfnamefont {M.~T.}\ \bibnamefont {Allen}},\ }\bibfield  {title} {\bibinfo {title} {Millikelvin microwave impedance microscopy in a dry dilution refrigerator},\ }\href {https://doi.org/10.1063/5.0159548} {\bibfield  {journal} {\bibinfo  {journal} {Review of Scientific Instruments}\ }\textbf {\bibinfo {volume} {94}},\ \bibinfo {pages} {093705} (\bibinfo {year} {2023})}\BibitemShut {NoStop}%
\bibitem [{\citenamefont {Griesmar}\ \emph {et~al.}(2021)\citenamefont {Griesmar}, \citenamefont {Rodriguez}, \citenamefont {Benzoni}, \citenamefont {Pillet}, \citenamefont {Smirr}, \citenamefont {Lafont},\ and\ \citenamefont {Girit}}]{Griesmar}%
  \BibitemOpen
  \bibfield  {author} {\bibinfo {author} {\bibfnamefont {J.}~\bibnamefont {Griesmar}}, \bibinfo {author} {\bibfnamefont {R.~H.}\ \bibnamefont {Rodriguez}}, \bibinfo {author} {\bibfnamefont {V.}~\bibnamefont {Benzoni}}, \bibinfo {author} {\bibfnamefont {J.-D.}\ \bibnamefont {Pillet}}, \bibinfo {author} {\bibfnamefont {J.-L.}\ \bibnamefont {Smirr}}, \bibinfo {author} {\bibfnamefont {F.}~\bibnamefont {Lafont}},\ and\ \bibinfo {author} {\bibfnamefont {{\c C}.~{\"O}.}\ \bibnamefont {Girit}},\ }\bibfield  {title} {\bibinfo {title} {Superconducting on-chip spectrometer for mesoscopic quantum systems},\ }\href {https://doi.org/10.1103/PhysRevResearch.3.043078} {\bibfield  {journal} {\bibinfo  {journal} {Physical Review Research}\ }\textbf {\bibinfo {volume} {3}},\ \bibinfo {pages} {043078} (\bibinfo {year} {2021})}\BibitemShut {NoStop}%
\bibitem [{\citenamefont {Silver}\ and\ \citenamefont {Zimmerman}(1967)}]{Silver1967}%
  \BibitemOpen
  \bibfield  {author} {\bibinfo {author} {\bibfnamefont {A.~H.}\ \bibnamefont {Silver}}\ and\ \bibinfo {author} {\bibfnamefont {J.~E.}\ \bibnamefont {Zimmerman}},\ }\bibfield  {title} {\bibinfo {title} {Multiple quantum resonance spectroscopy through weakly connected superconductors},\ }\href {https://doi.org/10.1063/1.1754885} {\bibfield  {journal} {\bibinfo  {journal} {Applied Physics Letters}\ }\textbf {\bibinfo {volume} {10}},\ \bibinfo {pages} {142} (\bibinfo {year} {1967})}\BibitemShut {NoStop}%
\bibitem [{\citenamefont {Bretheau}\ \emph {et~al.}(2013)\citenamefont {Bretheau}, \citenamefont {Girit}, \citenamefont {Pothier}, \citenamefont {Esteve},\ and\ \citenamefont {Urbina}}]{Bretheau2013}%
  \BibitemOpen
  \bibfield  {author} {\bibinfo {author} {\bibfnamefont {L.}~\bibnamefont {Bretheau}}, \bibinfo {author} {\bibfnamefont {{\c{C}}.~{\"O}.}\ \bibnamefont {Girit}}, \bibinfo {author} {\bibfnamefont {H.}~\bibnamefont {Pothier}}, \bibinfo {author} {\bibfnamefont {D.}~\bibnamefont {Esteve}},\ and\ \bibinfo {author} {\bibfnamefont {C.}~\bibnamefont {Urbina}},\ }\bibfield  {title} {\bibinfo {title} {Exciting andreev pairs in a superconducting atomic contact},\ }\href {https://doi.org/10.1038/nature12315} {\bibfield  {journal} {\bibinfo  {journal} {Nature}\ }\textbf {\bibinfo {volume} {499}},\ \bibinfo {pages} {312} (\bibinfo {year} {2013})}\BibitemShut {NoStop}%
\bibitem [{\citenamefont {Bretheau}\ \emph {et~al.}(2014)\citenamefont {Bretheau}, \citenamefont {Girit}, \citenamefont {Houzet}, \citenamefont {Pothier}, \citenamefont {Esteve},\ and\ \citenamefont {Urbina}}]{Girit2}%
  \BibitemOpen
  \bibfield  {author} {\bibinfo {author} {\bibfnamefont {L.}~\bibnamefont {Bretheau}}, \bibinfo {author} {\bibfnamefont {i.~m. c.~O.}\ \bibnamefont {Girit}}, \bibinfo {author} {\bibfnamefont {M.}~\bibnamefont {Houzet}}, \bibinfo {author} {\bibfnamefont {H.}~\bibnamefont {Pothier}}, \bibinfo {author} {\bibfnamefont {D.}~\bibnamefont {Esteve}},\ and\ \bibinfo {author} {\bibfnamefont {C.}~\bibnamefont {Urbina}},\ }\bibfield  {title} {\bibinfo {title} {Theory of microwave spectroscopy of andreev bound states with a josephson junction},\ }\href {https://doi.org/10.1103/PhysRevB.90.134506} {\bibfield  {journal} {\bibinfo  {journal} {Phys. Rev. B}\ }\textbf {\bibinfo {volume} {90}},\ \bibinfo {pages} {134506} (\bibinfo {year} {2014})}\BibitemShut {NoStop}%
\bibitem [{\citenamefont {Naaman}\ \emph {et~al.}(2001)\citenamefont {Naaman}, \citenamefont {Teizer},\ and\ \citenamefont {Dynes}}]{Dynes}%
  \BibitemOpen
  \bibfield  {author} {\bibinfo {author} {\bibfnamefont {O.}~\bibnamefont {Naaman}}, \bibinfo {author} {\bibfnamefont {W.}~\bibnamefont {Teizer}},\ and\ \bibinfo {author} {\bibfnamefont {R.~C.}\ \bibnamefont {Dynes}},\ }\bibfield  {title} {\bibinfo {title} {Fluctuation dominated josephson tunneling with a scanning tunneling microscope},\ }\href {https://doi.org/10.1103/PhysRevLett.87.097004} {\bibfield  {journal} {\bibinfo  {journal} {Phys. Rev. Lett.}\ }\textbf {\bibinfo {volume} {87}},\ \bibinfo {pages} {097004} (\bibinfo {year} {2001})}\BibitemShut {NoStop}%
\bibitem [{\citenamefont {Cho}\ \emph {et~al.}(2019)\citenamefont {Cho}, \citenamefont {Bastiaans}, \citenamefont {Chatzopoulos}, \citenamefont {Gu},\ and\ \citenamefont {Allan}}]{Cho2019}%
  \BibitemOpen
  \bibfield  {author} {\bibinfo {author} {\bibfnamefont {D.}~\bibnamefont {Cho}}, \bibinfo {author} {\bibfnamefont {K.~M.}\ \bibnamefont {Bastiaans}}, \bibinfo {author} {\bibfnamefont {D.}~\bibnamefont {Chatzopoulos}}, \bibinfo {author} {\bibfnamefont {G.~D.}\ \bibnamefont {Gu}},\ and\ \bibinfo {author} {\bibfnamefont {M.~P.}\ \bibnamefont {Allan}},\ }\bibfield  {title} {\bibinfo {title} {A strongly inhomogeneous superfluid in an iron-based superconductor},\ }\href {https://doi.org/10.1038/s41586-019-1408-8} {\bibfield  {journal} {\bibinfo  {journal} {Nature}\ }\textbf {\bibinfo {volume} {571}},\ \bibinfo {pages} {541} (\bibinfo {year} {2019})}\BibitemShut {NoStop}%
\bibitem [{\citenamefont {Hamidian}\ \emph {et~al.}(2016)\citenamefont {Hamidian}, \citenamefont {Edkins}, \citenamefont {Joo}, \citenamefont {Kostin}, \citenamefont {Eisaki}, \citenamefont {Uchida}, \citenamefont {Lawler}, \citenamefont {Kim}, \citenamefont {Mackenzie}, \citenamefont {Fujita}, \citenamefont {Lee},\ and\ \citenamefont {Davis}}]{Hamidian2016}%
  \BibitemOpen
  \bibfield  {author} {\bibinfo {author} {\bibfnamefont {M.~H.}\ \bibnamefont {Hamidian}}, \bibinfo {author} {\bibfnamefont {S.~D.}\ \bibnamefont {Edkins}}, \bibinfo {author} {\bibfnamefont {S.~H.}\ \bibnamefont {Joo}}, \bibinfo {author} {\bibfnamefont {A.}~\bibnamefont {Kostin}}, \bibinfo {author} {\bibfnamefont {H.}~\bibnamefont {Eisaki}}, \bibinfo {author} {\bibfnamefont {S.}~\bibnamefont {Uchida}}, \bibinfo {author} {\bibfnamefont {M.~J.}\ \bibnamefont {Lawler}}, \bibinfo {author} {\bibfnamefont {E.-A.}\ \bibnamefont {Kim}}, \bibinfo {author} {\bibfnamefont {A.~P.}\ \bibnamefont {Mackenzie}}, \bibinfo {author} {\bibfnamefont {K.}~\bibnamefont {Fujita}}, \bibinfo {author} {\bibfnamefont {J.}~\bibnamefont {Lee}},\ and\ \bibinfo {author} {\bibfnamefont {J.~C.~S.}\ \bibnamefont {Davis}},\ }\bibfield  {title} {\bibinfo {title} {Detection of a cooper-pair density wave in bi2sr2cacu2o8+x},\ }\href {https://doi.org/10.1038/nature17411} {\bibfield  {journal} {\bibinfo  {journal} {Nature}\ }\textbf {\bibinfo
  {volume} {532}},\ \bibinfo {pages} {343} (\bibinfo {year} {2016})}\BibitemShut {NoStop}%
\bibitem [{\citenamefont {Randeria}\ \emph {et~al.}(2016)\citenamefont {Randeria}, \citenamefont {Feldman}, \citenamefont {Drozdov},\ and\ \citenamefont {Yazdani}}]{randeria2016}%
  \BibitemOpen
  \bibfield  {author} {\bibinfo {author} {\bibfnamefont {M.~T.}\ \bibnamefont {Randeria}}, \bibinfo {author} {\bibfnamefont {B.~E.}\ \bibnamefont {Feldman}}, \bibinfo {author} {\bibfnamefont {I.~K.}\ \bibnamefont {Drozdov}},\ and\ \bibinfo {author} {\bibfnamefont {A.}~\bibnamefont {Yazdani}},\ }\bibfield  {title} {\bibinfo {title} {Scanning josephson spectroscopy on the atomic scale},\ }\href@noop {} {\bibfield  {journal} {\bibinfo  {journal} {Physical Review B}\ }\textbf {\bibinfo {volume} {93}},\ \bibinfo {pages} {161115} (\bibinfo {year} {2016})}\BibitemShut {NoStop}%
\bibitem [{\citenamefont {Proslier}\ \emph {et~al.}(2006)\citenamefont {Proslier}, \citenamefont {Kohen}, \citenamefont {Noat}, \citenamefont {Cren}, \citenamefont {Roditchev},\ and\ \citenamefont {Sacks}}]{proslier2006}%
  \BibitemOpen
  \bibfield  {author} {\bibinfo {author} {\bibfnamefont {T.}~\bibnamefont {Proslier}}, \bibinfo {author} {\bibfnamefont {A.}~\bibnamefont {Kohen}}, \bibinfo {author} {\bibfnamefont {Y.}~\bibnamefont {Noat}}, \bibinfo {author} {\bibfnamefont {T.}~\bibnamefont {Cren}}, \bibinfo {author} {\bibfnamefont {D.}~\bibnamefont {Roditchev}},\ and\ \bibinfo {author} {\bibfnamefont {W.}~\bibnamefont {Sacks}},\ }\bibfield  {title} {\bibinfo {title} {Probing the superconducting condensate on a nanometer scale},\ }\href@noop {} {\bibfield  {journal} {\bibinfo  {journal} {Europhysics Letters}\ }\textbf {\bibinfo {volume} {73}},\ \bibinfo {pages} {962} (\bibinfo {year} {2006})}\BibitemShut {NoStop}%
\bibitem [{\citenamefont {Rodrigo}\ \emph {et~al.}(2006)\citenamefont {Rodrigo}, \citenamefont {Crespo},\ and\ \citenamefont {Vieira}}]{rodrigo2006}%
  \BibitemOpen
  \bibfield  {author} {\bibinfo {author} {\bibfnamefont {J.}~\bibnamefont {Rodrigo}}, \bibinfo {author} {\bibfnamefont {V.}~\bibnamefont {Crespo}},\ and\ \bibinfo {author} {\bibfnamefont {S.}~\bibnamefont {Vieira}},\ }\bibfield  {title} {\bibinfo {title} {Josephson current at atomic scale: tunneling and nanocontacts using a stm},\ }\href@noop {} {\bibfield  {journal} {\bibinfo  {journal} {Physica C: Superconductivity and its applications}\ }\textbf {\bibinfo {volume} {437}},\ \bibinfo {pages} {270} (\bibinfo {year} {2006})}\BibitemShut {NoStop}%
\bibitem [{\citenamefont {Bergeal}\ \emph {et~al.}(2008)\citenamefont {Bergeal}, \citenamefont {Noat}, \citenamefont {Cren}, \citenamefont {Proslier}, \citenamefont {Dubost}, \citenamefont {Debontridder}, \citenamefont {Zimmers}, \citenamefont {Roditchev}, \citenamefont {Sacks},\ and\ \citenamefont {Marcus}}]{bergeal2008}%
  \BibitemOpen
  \bibfield  {author} {\bibinfo {author} {\bibfnamefont {N.}~\bibnamefont {Bergeal}}, \bibinfo {author} {\bibfnamefont {Y.}~\bibnamefont {Noat}}, \bibinfo {author} {\bibfnamefont {T.}~\bibnamefont {Cren}}, \bibinfo {author} {\bibfnamefont {T.}~\bibnamefont {Proslier}}, \bibinfo {author} {\bibfnamefont {V.}~\bibnamefont {Dubost}}, \bibinfo {author} {\bibfnamefont {F.}~\bibnamefont {Debontridder}}, \bibinfo {author} {\bibfnamefont {A.}~\bibnamefont {Zimmers}}, \bibinfo {author} {\bibfnamefont {D.}~\bibnamefont {Roditchev}}, \bibinfo {author} {\bibfnamefont {W.}~\bibnamefont {Sacks}},\ and\ \bibinfo {author} {\bibfnamefont {J.}~\bibnamefont {Marcus}},\ }\bibfield  {title} {\bibinfo {title} {Mapping the superconducting condensate surrounding a vortex in superconducting v 3 si using a superconducting mgb 2 tip in a scanning tunneling microscope},\ }\href@noop {} {\bibfield  {journal} {\bibinfo  {journal} {Physical Review B}\ }\textbf {\bibinfo {volume} {78}},\ \bibinfo {pages} {140507} (\bibinfo {year}
  {2008})}\BibitemShut {NoStop}%
\bibitem [{\citenamefont {Pan}\ \emph {et~al.}(1998)\citenamefont {Pan}, \citenamefont {Hudson},\ and\ \citenamefont {Davis}}]{Pan1998}%
  \BibitemOpen
  \bibfield  {author} {\bibinfo {author} {\bibfnamefont {S.~H.}\ \bibnamefont {Pan}}, \bibinfo {author} {\bibfnamefont {E.~W.}\ \bibnamefont {Hudson}},\ and\ \bibinfo {author} {\bibfnamefont {J.~C.}\ \bibnamefont {Davis}},\ }\bibfield  {title} {\bibinfo {title} {Vacuum tunneling of superconducting quasiparticles from atomically sharp scanning tunneling microscope tips},\ }\href {https://doi.org/10.1063/1.122654} {\bibfield  {journal} {\bibinfo  {journal} {Applied Physics Letters}\ }\textbf {\bibinfo {volume} {73}},\ \bibinfo {pages} {2992} (\bibinfo {year} {1998})}\BibitemShut {NoStop}%
\bibitem [{\citenamefont {Franke}\ \emph {et~al.}(2011)\citenamefont {Franke}, \citenamefont {Schulze},\ and\ \citenamefont {Pascual}}]{Franke2011}%
  \BibitemOpen
  \bibfield  {author} {\bibinfo {author} {\bibfnamefont {K.~J.}\ \bibnamefont {Franke}}, \bibinfo {author} {\bibfnamefont {G.}~\bibnamefont {Schulze}},\ and\ \bibinfo {author} {\bibfnamefont {J.~I.}\ \bibnamefont {Pascual}},\ }\bibfield  {title} {\bibinfo {title} {Competition of superconducting phenomena and kondo screening at the nanoscale},\ }\href {https://doi.org/10.1126/science.1202204} {\bibfield  {journal} {\bibinfo  {journal} {Science}\ }\textbf {\bibinfo {volume} {332}},\ \bibinfo {pages} {940} (\bibinfo {year} {2011})}\BibitemShut {NoStop}%
\bibitem [{\citenamefont {Ruby}\ \emph {et~al.}(2015)\citenamefont {Ruby}, \citenamefont {Pientka}, \citenamefont {Peng}, \citenamefont {von Oppen}, \citenamefont {Heinrich},\ and\ \citenamefont {Franke}}]{Franke2015}%
  \BibitemOpen
  \bibfield  {author} {\bibinfo {author} {\bibfnamefont {M.}~\bibnamefont {Ruby}}, \bibinfo {author} {\bibfnamefont {F.}~\bibnamefont {Pientka}}, \bibinfo {author} {\bibfnamefont {Y.}~\bibnamefont {Peng}}, \bibinfo {author} {\bibfnamefont {F.}~\bibnamefont {von Oppen}}, \bibinfo {author} {\bibfnamefont {B.~W.}\ \bibnamefont {Heinrich}},\ and\ \bibinfo {author} {\bibfnamefont {K.~J.}\ \bibnamefont {Franke}},\ }\bibfield  {title} {\bibinfo {title} {Tunneling processes into localized subgap states in superconductors},\ }\href {https://doi.org/10.1103/PhysRevLett.115.087001} {\bibfield  {journal} {\bibinfo  {journal} {Phys. Rev. Lett.}\ }\textbf {\bibinfo {volume} {115}},\ \bibinfo {pages} {087001} (\bibinfo {year} {2015})}\BibitemShut {NoStop}%
\bibitem [{\citenamefont {J{\"a}ck}\ \emph {et~al.}(2015)\citenamefont {J{\"a}ck}, \citenamefont {Eltschka}, \citenamefont {Assig}, \citenamefont {Hardock}, \citenamefont {Etzkorn}, \citenamefont {Ast},\ and\ \citenamefont {Kern}}]{Jack2015}%
  \BibitemOpen
  \bibfield  {author} {\bibinfo {author} {\bibfnamefont {B.}~\bibnamefont {J{\"a}ck}}, \bibinfo {author} {\bibfnamefont {M.}~\bibnamefont {Eltschka}}, \bibinfo {author} {\bibfnamefont {M.}~\bibnamefont {Assig}}, \bibinfo {author} {\bibfnamefont {A.}~\bibnamefont {Hardock}}, \bibinfo {author} {\bibfnamefont {M.}~\bibnamefont {Etzkorn}}, \bibinfo {author} {\bibfnamefont {C.~R.}\ \bibnamefont {Ast}},\ and\ \bibinfo {author} {\bibfnamefont {K.}~\bibnamefont {Kern}},\ }\bibfield  {title} {\bibinfo {title} {A nanoscale gigahertz source realized with josephson scanning tunneling microscopy},\ }\href {https://doi.org/10.1063/1.4905322} {\bibfield  {journal} {\bibinfo  {journal} {Applied Physics Letters}\ }\textbf {\bibinfo {volume} {106}},\ \bibinfo {pages} {013109} (\bibinfo {year} {2015})}\BibitemShut {NoStop}%
\bibitem [{\citenamefont {Ast}\ \emph {et~al.}(2016)\citenamefont {Ast}, \citenamefont {J{\"a}ck}, \citenamefont {Senkpiel}, \citenamefont {Eltschka}, \citenamefont {Etzkorn}, \citenamefont {Ankerhold},\ and\ \citenamefont {Kern}}]{Ast2016}%
  \BibitemOpen
  \bibfield  {author} {\bibinfo {author} {\bibfnamefont {C.~R.}\ \bibnamefont {Ast}}, \bibinfo {author} {\bibfnamefont {B.}~\bibnamefont {J{\"a}ck}}, \bibinfo {author} {\bibfnamefont {J.}~\bibnamefont {Senkpiel}}, \bibinfo {author} {\bibfnamefont {M.}~\bibnamefont {Eltschka}}, \bibinfo {author} {\bibfnamefont {M.}~\bibnamefont {Etzkorn}}, \bibinfo {author} {\bibfnamefont {J.}~\bibnamefont {Ankerhold}},\ and\ \bibinfo {author} {\bibfnamefont {K.}~\bibnamefont {Kern}},\ }\bibfield  {title} {\bibinfo {title} {Sensing the quantum limit in scanning tunnelling spectroscopy},\ }\href {https://doi.org/10.1038/ncomms13009} {\bibfield  {journal} {\bibinfo  {journal} {Nature Communications}\ }\textbf {\bibinfo {volume} {7}},\ \bibinfo {pages} {13009} (\bibinfo {year} {2016})}\BibitemShut {NoStop}%
\bibitem [{\citenamefont {Roychowdhury}\ \emph {et~al.}(2015)\citenamefont {Roychowdhury}, \citenamefont {Dreyer}, \citenamefont {Anderson}, \citenamefont {Lobb},\ and\ \citenamefont {Wellstood}}]{roychowdhury2015}%
  \BibitemOpen
  \bibfield  {author} {\bibinfo {author} {\bibfnamefont {A.}~\bibnamefont {Roychowdhury}}, \bibinfo {author} {\bibfnamefont {M.}~\bibnamefont {Dreyer}}, \bibinfo {author} {\bibfnamefont {J.}~\bibnamefont {Anderson}}, \bibinfo {author} {\bibfnamefont {C.}~\bibnamefont {Lobb}},\ and\ \bibinfo {author} {\bibfnamefont {F.}~\bibnamefont {Wellstood}},\ }\bibfield  {title} {\bibinfo {title} {Microwave photon-assisted incoherent cooper-pair tunneling in a josephson stm},\ }\href@noop {} {\bibfield  {journal} {\bibinfo  {journal} {Physical Review Applied}\ }\textbf {\bibinfo {volume} {4}},\ \bibinfo {pages} {034011} (\bibinfo {year} {2015})}\BibitemShut {NoStop}%
\bibitem [{\citenamefont {Bastiaans}\ \emph {et~al.}(2019)\citenamefont {Bastiaans}, \citenamefont {Cho}, \citenamefont {Chatzopoulos}, \citenamefont {Leeuwenhoek}, \citenamefont {Koks},\ and\ \citenamefont {Allan}}]{bastiaans2019}%
  \BibitemOpen
  \bibfield  {author} {\bibinfo {author} {\bibfnamefont {K.}~\bibnamefont {Bastiaans}}, \bibinfo {author} {\bibfnamefont {D.}~\bibnamefont {Cho}}, \bibinfo {author} {\bibfnamefont {D.}~\bibnamefont {Chatzopoulos}}, \bibinfo {author} {\bibfnamefont {M.}~\bibnamefont {Leeuwenhoek}}, \bibinfo {author} {\bibfnamefont {C.}~\bibnamefont {Koks}},\ and\ \bibinfo {author} {\bibfnamefont {M.}~\bibnamefont {Allan}},\ }\bibfield  {title} {\bibinfo {title} {Imaging doubled shot noise in a josephson scanning tunneling microscope},\ }\href@noop {} {\bibfield  {journal} {\bibinfo  {journal} {Physical Review B}\ }\textbf {\bibinfo {volume} {100}},\ \bibinfo {pages} {104506} (\bibinfo {year} {2019})}\BibitemShut {NoStop}%
\bibitem [{\citenamefont {Pekola}\ \emph {et~al.}(2010)\citenamefont {Pekola}, \citenamefont {Maisi}, \citenamefont {Kafanov}, \citenamefont {Chekurov}, \citenamefont {Kemppinen}, \citenamefont {Pashkin}, \citenamefont {Saira}, \citenamefont {M\"ott\"onen},\ and\ \citenamefont {Tsai}}]{EnvironmentAssisted2010}%
  \BibitemOpen
  \bibfield  {author} {\bibinfo {author} {\bibfnamefont {J.~P.}\ \bibnamefont {Pekola}}, \bibinfo {author} {\bibfnamefont {V.~F.}\ \bibnamefont {Maisi}}, \bibinfo {author} {\bibfnamefont {S.}~\bibnamefont {Kafanov}}, \bibinfo {author} {\bibfnamefont {N.}~\bibnamefont {Chekurov}}, \bibinfo {author} {\bibfnamefont {A.}~\bibnamefont {Kemppinen}}, \bibinfo {author} {\bibfnamefont {Y.~A.}\ \bibnamefont {Pashkin}}, \bibinfo {author} {\bibfnamefont {O.-P.}\ \bibnamefont {Saira}}, \bibinfo {author} {\bibfnamefont {M.}~\bibnamefont {M\"ott\"onen}},\ and\ \bibinfo {author} {\bibfnamefont {J.~S.}\ \bibnamefont {Tsai}},\ }\bibfield  {title} {\bibinfo {title} {Environment-assisted tunneling as an origin of the dynes density of states},\ }\href {https://doi.org/10.1103/PhysRevLett.105.026803} {\bibfield  {journal} {\bibinfo  {journal} {Phys. Rev. Lett.}\ }\textbf {\bibinfo {volume} {105}},\ \bibinfo {pages} {026803} (\bibinfo {year} {2010})}\BibitemShut {NoStop}%
\bibitem [{\citenamefont {Likharev}(1985)}]{likharev1985}%
  \BibitemOpen
  \bibfield  {author} {\bibinfo {author} {\bibfnamefont {K.}~\bibnamefont {Likharev}},\ }\bibfield  {title} {\bibinfo {title} {Introduction to the dynamics of josephson junctions},\ }\href@noop {} {\bibfield  {journal} {\bibinfo  {journal} {Moscow Izdatel Nauka}\ } (\bibinfo {year} {1985})}\BibitemShut {NoStop}%
\bibitem [{\citenamefont {Noat}\ \emph {et~al.}(2013)\citenamefont {Noat}, \citenamefont {Cherkez}, \citenamefont {Brun}, \citenamefont {Cren}, \citenamefont {Carbillet}, \citenamefont {Debontridder}, \citenamefont {Ilin}, \citenamefont {Siegel}, \citenamefont {Semenov}, \citenamefont {H\"ubers},\ and\ \citenamefont {Roditchev}}]{Brun}%
  \BibitemOpen
  \bibfield  {author} {\bibinfo {author} {\bibfnamefont {Y.}~\bibnamefont {Noat}}, \bibinfo {author} {\bibfnamefont {V.}~\bibnamefont {Cherkez}}, \bibinfo {author} {\bibfnamefont {C.}~\bibnamefont {Brun}}, \bibinfo {author} {\bibfnamefont {T.}~\bibnamefont {Cren}}, \bibinfo {author} {\bibfnamefont {C.}~\bibnamefont {Carbillet}}, \bibinfo {author} {\bibfnamefont {F.}~\bibnamefont {Debontridder}}, \bibinfo {author} {\bibfnamefont {K.}~\bibnamefont {Ilin}}, \bibinfo {author} {\bibfnamefont {M.}~\bibnamefont {Siegel}}, \bibinfo {author} {\bibfnamefont {A.}~\bibnamefont {Semenov}}, \bibinfo {author} {\bibfnamefont {H.-W.}\ \bibnamefont {H\"ubers}},\ and\ \bibinfo {author} {\bibfnamefont {D.}~\bibnamefont {Roditchev}},\ }\bibfield  {title} {\bibinfo {title} {Unconventional superconductivity in ultrathin superconducting nbn films studied by scanning tunneling spectroscopy},\ }\href {https://doi.org/10.1103/PhysRevB.88.014503} {\bibfield  {journal} {\bibinfo  {journal} {Phys. Rev. B}\ }\textbf {\bibinfo {volume}
  {88}},\ \bibinfo {pages} {014503} (\bibinfo {year} {2013})}\BibitemShut {NoStop}%
\bibitem [{\citenamefont {Devoret}\ \emph {et~al.}(1990)\citenamefont {Devoret}, \citenamefont {Esteve}, \citenamefont {Grabert}, \citenamefont {Ingold}, \citenamefont {Pothier},\ and\ \citenamefont {Urbina}}]{Devoret}%
  \BibitemOpen
  \bibfield  {author} {\bibinfo {author} {\bibfnamefont {M.~H.}\ \bibnamefont {Devoret}}, \bibinfo {author} {\bibfnamefont {D.}~\bibnamefont {Esteve}}, \bibinfo {author} {\bibfnamefont {H.}~\bibnamefont {Grabert}}, \bibinfo {author} {\bibfnamefont {G.-L.}\ \bibnamefont {Ingold}}, \bibinfo {author} {\bibfnamefont {H.}~\bibnamefont {Pothier}},\ and\ \bibinfo {author} {\bibfnamefont {C.}~\bibnamefont {Urbina}},\ }\bibfield  {title} {\bibinfo {title} {Effect of the electromagnetic environment on the coulomb blockade in ultrasmall tunnel junctions},\ }\href {https://doi.org/10.1103/PhysRevLett.64.1824} {\bibfield  {journal} {\bibinfo  {journal} {Phys. Rev. Lett.}\ }\textbf {\bibinfo {volume} {64}},\ \bibinfo {pages} {1824} (\bibinfo {year} {1990})}\BibitemShut {NoStop}%
\bibitem [{\citenamefont {Dynes}\ \emph {et~al.}(1978)\citenamefont {Dynes}, \citenamefont {Narayanamurti},\ and\ \citenamefont {Garno}}]{DynesEq}%
  \BibitemOpen
  \bibfield  {author} {\bibinfo {author} {\bibfnamefont {R.~C.}\ \bibnamefont {Dynes}}, \bibinfo {author} {\bibfnamefont {V.}~\bibnamefont {Narayanamurti}},\ and\ \bibinfo {author} {\bibfnamefont {J.~P.}\ \bibnamefont {Garno}},\ }\bibfield  {title} {\bibinfo {title} {Direct measurement of quasiparticle-lifetime broadening in a strong-coupled superconductor},\ }\href {https://doi.org/10.1103/PhysRevLett.41.1509} {\bibfield  {journal} {\bibinfo  {journal} {Phys. Rev. Lett.}\ }\textbf {\bibinfo {volume} {41}},\ \bibinfo {pages} {1509} (\bibinfo {year} {1978})}\BibitemShut {NoStop}%
\bibitem [{\citenamefont {Ingold}\ and\ \citenamefont {Nazarov}(1992)}]{Ingold1992}%
  \BibitemOpen
  \bibfield  {author} {\bibinfo {author} {\bibfnamefont {G.-L.}\ \bibnamefont {Ingold}}\ and\ \bibinfo {author} {\bibfnamefont {Y.~V.}\ \bibnamefont {Nazarov}},\ }\bibinfo {title} {Charge tunneling rates in ultrasmall junctions},\ in\ \href {https://doi.org/10.1007/978-1-4757-2166-9_2} {\emph {\bibinfo {booktitle} {Single Charge Tunneling: Coulomb Blockade Phenomena In Nanostructures}}},\ \bibinfo {editor} {edited by\ \bibinfo {editor} {\bibfnamefont {H.}~\bibnamefont {Grabert}}\ and\ \bibinfo {editor} {\bibfnamefont {M.~H.}\ \bibnamefont {Devoret}}}\ (\bibinfo  {publisher} {Springer US},\ \bibinfo {address} {Boston, MA},\ \bibinfo {year} {1992})\ pp.\ \bibinfo {pages} {21--107}\BibitemShut {NoStop}%
\bibitem [{\citenamefont {Semenov}\ \emph {et~al.}(2009)\citenamefont {Semenov}, \citenamefont {G\"unther}, \citenamefont {B\"ottger}, \citenamefont {H\"ubers}, \citenamefont {Bartolf}, \citenamefont {Engel}, \citenamefont {Schilling}, \citenamefont {Ilin}, \citenamefont {Siegel}, \citenamefont {Schneider}, \citenamefont {Gerthsen},\ and\ \citenamefont {Gippius}}]{nbnlayer}%
  \BibitemOpen
  \bibfield  {author} {\bibinfo {author} {\bibfnamefont {A.}~\bibnamefont {Semenov}}, \bibinfo {author} {\bibfnamefont {B.}~\bibnamefont {G\"unther}}, \bibinfo {author} {\bibfnamefont {U.}~\bibnamefont {B\"ottger}}, \bibinfo {author} {\bibfnamefont {H.-W.}\ \bibnamefont {H\"ubers}}, \bibinfo {author} {\bibfnamefont {H.}~\bibnamefont {Bartolf}}, \bibinfo {author} {\bibfnamefont {A.}~\bibnamefont {Engel}}, \bibinfo {author} {\bibfnamefont {A.}~\bibnamefont {Schilling}}, \bibinfo {author} {\bibfnamefont {K.}~\bibnamefont {Ilin}}, \bibinfo {author} {\bibfnamefont {M.}~\bibnamefont {Siegel}}, \bibinfo {author} {\bibfnamefont {R.}~\bibnamefont {Schneider}}, \bibinfo {author} {\bibfnamefont {D.}~\bibnamefont {Gerthsen}},\ and\ \bibinfo {author} {\bibfnamefont {N.~A.}\ \bibnamefont {Gippius}},\ }\bibfield  {title} {\bibinfo {title} {Optical and transport properties of ultrathin nbn films and nanostructures},\ }\href {https://doi.org/10.1103/PhysRevB.80.054510} {\bibfield  {journal} {\bibinfo  {journal} {Phys. Rev.
  B}\ }\textbf {\bibinfo {volume} {80}},\ \bibinfo {pages} {054510} (\bibinfo {year} {2009})}\BibitemShut {NoStop}%
\end{thebibliography}%

\end{document}